\newif\ifnotjournalversion

\notjournalversiontrue
\ifnotjournalversion 
\documentclass[11pt]{article}
\usepackage{times,fullpage,amsmath,amsthm,amssymb,hyperref,graphicx,color,subcaption}

\title{Hopcroft's Problem, Log-Star Shaving, 2D Fractional Cascading, and Decision Trees%
\footnote{A preliminary version of this paper has appeared in \emph{Proc. 33rd ACM-SIAM Symposium on Discrete Algorithms (SODA)}, pages 190--210, 2022.}}

\author{Timothy M. Chan%
  \thanks{Department of Computer Science,
    University of Illinois at Urbana-Champaign,
    \{tmc,dwzheng2\}@illinois.edu.  Work supported by
    NSF Grant CCF-1814026.} 
  \and
  Da Wei Zheng\textsuperscript{\textdagger}}

\else
\documentclass[manuscript]{acmart}
\usepackage{times,amsmath,amsthm,graphicx,color,subcaption}

\AtBeginDocument{%
  \providecommand\BibTeX{{%
    \normalfont B\kern-0.5em{\scshape i\kern-0.25em b}\kern-0.8em\TeX}}}

\setcopyright{acmcopyright}
\copyrightyear{2018}
\acmYear{2018}
\acmDOI{XXXXXXX.XXXXXXX}

\acmPrice{15.00}
\acmISBN{978-1-4503-XXXX-X/18/06}

\title{Hopcroft's Problem, Log-Star Shaving, 2D Fractional Cascading, and Decision Trees}
\titlenote{A preliminary version of this paper has appeared in \emph{Proc. 33rd ACM-SIAM Symposium on Discrete Algorithms (SODA)}, pages 190--210, 2022.}

\author{Timothy M. Chan}
\authornote{Work supported by NSF Grant CCF-1814026.}
\email{tmc@illinois.edu}
\orcid{0000-0002-8093-0675}
\affiliation{%
  \institution{University of Illinois at Urbana-Champaign}
  \streetaddress{201 N. Goodwin Ave.}
  \city{Urbana}
  \state{Illinois}
  \country{USA}
  \postcode{61801}
}

\author{Da Wei Zheng}
\email{dwzheng2@illinois.edu}
\orcid{0000-0002-0844-9457}
\affiliation{%
  \institution{University of Illinois at Urbana-Champaign}
  \streetaddress{201 N. Goodwin Ave.}
  \city{Urbana}
  \state{Illinois}
  \country{USA}
  \postcode{61801}
}

\fi

\newcommand{\eps}{\varepsilon}
\newcommand{\D}{\Delta}
\newcommand{\R}{\mathbb{R}}
\newcommand{\E}{\mathbb{E}}

\newcommand{\up}[1]{\left\lceil#1\right\rceil}
\newcommand{\down}[1]{\left\lfloor#1\right\rfloor}
\newcommand{\IGNORE}[1]{}
\newcommand{\A}{\mathcal{A}}
\newcommand{\VD}{\mathcal{VD}}%
\newtheorem{theorem}{Theorem}[section]
\newtheorem{lemma}[theorem]{Lemma}
\newtheorem{corollary}[theorem]{Corollary}
\newtheorem{problem}[theorem]{Problem}
\newenvironment{myproof}{\begin{proof}}{\end{proof}}

\begin{document}

\maketitle
\begin{abstract}
We revisit Hopcroft's problem and related fundamental problems about geometric range searching. Given $n$ points and $n$ lines in the plane, we show how to count the number of point-line incidence pairs or the number of point-above-line pairs in $O(n^{4/3})$ time, which matches the conjectured lower bound and improves the best previous time bound of $n^{4/3}2^{O(\log^*n)}$ obtained almost 30 years ago by Matou\v sek.

We describe two interesting and different ways to achieve the result: the first is randomized and uses a new 2D version of fractional cascading for arrangements of lines; the second is deterministic and uses decision trees in a manner inspired by the sorting technique of Fredman (1976).  The second approach extends to any constant dimension.

Many consequences follow from these new ideas: for example, we obtain an $O(n^{4/3})$-time algorithm for line segment intersection counting in the plane, $O(n^{4/3})$-time randomized algorithms for distance selection in the plane and bichromatic closest pair and Euclidean minimum spanning tree in three or four dimensions, and a randomized data structure for halfplane range counting in the plane with $O(n^{4/3})$ preprocessing time and space and $O(n^{1/3})$ query time.
\end{abstract}
\section{Introduction}

In the 1980s and 1990s, many low-dimensional
problems in computational geometry were
discovered to have polynomial-time algorithms with
fractional exponents in their time complexities.
For example, Yao~\cite{Yao82}  in 1982 described the first
subquadratic algorithm for computing
the Euclidean minimum spanning tree of $n$ points
in any constant dimension $d$,
with running time near $n^{2-1/2^{d+1}}$; in the 3D case, the 
time bound was near $n^{9/5}$.
Subsequently, the exponents were improved; in particular, in the 3D case, the best time bound was near $n^{4/3}$~\cite{AgarwalES91}.
For another example, Chazelle~\cite{Chazelle86}  in 1986  described 
the first subquadratic algorithm for counting the number of intersections 
of $n$ line segments in 2D,
with running time $O(n^{1.695})$.  Eventually, the time bound was improved to near $n^{4/3}$ \cite{GuibasOS88,Agarwal90ii,Chazelle93}.
Other examples abound in the literature, too numerous to be listed here.
All these problems are intimately related to range searching~\cite{AgarwalE99,BergCKO08}, and often the
exponents arise  (implicitly or otherwise) from balancing the preprocessing
cost and the cost of answering multiple range searching queries.

After years of research, we have now reached what are conjectured
to be the optimal exponents for many fundamental geometric problems. 
For example, $n^{4/3}$ is likely tight for 
segment intersection counting: although unconditional
lower bounds in general models of computation are difficult to establish, 
Erickson~\cite{Erickson96,Erickson95} proved $\Omega(n^{4/3})$ lower bounds 
for this and other related problems in 
a restricted model which he called ``partitioning algorithms''. 
Erickson's lower bound applied to an even more
basic problem, \emph{Hopcroft's problem}\footnote{
First posed by John Hopcroft in the early 1980s.
}: given $n$ points and $n$ lines in 2D, 
detect (or count, or report all) point-line incidence pairs.\footnote{
Note that the number of incidences is known to be $\Theta(n^{4/3})$ in
the worst case---this is the well-known Sz\'emeredi--Trotter Theorem~\cite{SzemerediT83},
with lower bound construction provided by Erd\"os.  So, $\Omega(n^{4/3})$ is a trivial lower bound on the time complexity for the
``report all'' version of Hopcroft's problem.
}
Furthermore, for some closely related range
searching problems with weights, Chazelle \cite{Chazelle89,Chazelle97} proved near-$n^{4/3}$ 
lower bounds in the arithmetic semigroup model.

Despite the successes in this impressive body of work in computational geometry,
a blemish remains: All these algorithms with fractional exponents
have extra factors in their time bounds, usually of the form $n^{\eps}$ for an arbitrarily small constant $\eps>0$, or $\log^{O(1)}n$ for an undetermined number of logarithmic
factors.  
In a few cases (see below), iterated logarithmic factors like $2^{O(\log^*n)}$ turn up, surprisingly.

For example, for Hopcroft's problem and segment intersection counting,
an $O(n^{4/3+\eps})$-time (randomized) algorithm was obtained by Edelsbrunner, Guibas, and Sharir~\cite{EdelsbrunnerGS90} and
Guibas, Overmars, and Sharir~\cite{GuibasOS88}, before
Agarwal~\cite{Agarwal90ii} improved the running time
to $O(n^{4/3}\log^{1.78} n)$,
and then Chazelle~\cite{Chazelle93} to $O(n^{4/3}\log^{1/3}n)$.
For Hopcroft's problem, Matou\v sek~\cite{Matousek93}   in 1993 managed to
further reduce the bound to $n^{4/3}2^{O(\log^*n)}$.

\newcommand{\Paragraph}[1]{\paragraph{#1}}
\Paragraph{Main new result.}
We show that Hopcroft's problem in 2D can be solved in
$O(n^{4/3})$ time, without any extra factors!  This improves Matou\v sek's
result from almost 30 years ago.

\Paragraph{Significance.}
Although counting incidences in 2D may not sound very useful  by itself, Hopcroft's problem is significant because it is the simplest representative for a whole class of problems about nonorthogonal range searching.  It reduces to offline range searching, namely, answering a batch of $n$ range queries on $n$ points, where the ranges are lines.
Our results generalize to Hopcroft's problem in any constant dimension $d$
(detecting/counting incidence pairs for $n$ points and $n$ hyperplanes
in $O(n^{2d/(d+1)})$ time) and to
different variants (for example, counting point-above-hyperplane pairs, or offline halfspace range counting),
and indeed our ideas yield
new algorithms for a long list of problems in this domain, including segment intersection counting, and Euclidean minimum spanning trees in any constant
dimensions, as well as new data structures for online 2D halfplane range counting queries 
(see below for the statements of these results).

While shaving an iterated logarithmic factor may seem minor from the
practical perspective, the value of our work is in
cleaning up existing theoretical bounds
for a central class of problems in computational geometry.

\Paragraph{Difficulty of $\log^*$ shaving.}
The presence of some logarithmic factors might seem inevitable for this type of problem: for example,
in data structures for 2D halfplane range counting,
under standard comparison-based models, 
query time
must be $\Omega(\log n)$ even if we allow large preprocessing time, and 
preprocessing time must be $\Omega(n\log n)$ even if we allow large ($o(n^{1-\eps})$)
query time.  The known near-$n^{4/3}$ solutions to Hopcroft's problem were essentially obtained by interpolating between these two extremes.  With a smart recursion, Matou\v sek~\cite{Matousek93} managed to lower the effect of the logarithmic factor to iterated logarithm, but recursion alone cannot
get rid of the extra factors entirely, and some new ideas are needed.\footnote{
At the end of his paper~\cite{Matousek93}, Matou\v sek wrote about the challenge in making further improvements: ``\ldots\ the $2^{O(\log^*n)}$ factor
originates in the following manner: we are unable
to solve the problem in a constant number of stages
with the present method, essentialy [sic] because
a nonconstant time is spent for location of every point
in the cutting.  Every stage contributes a constant
multiplicative factor to the `excess' in the number
of subproblems.  This is because we lack some mechanism
to control this, similar to Chazelle's method (he can use
the number of intersections of the lines as the control
device, but we flip the roles of lines
and points at every stage, and such a control device is
missing in this situation). 

``[\ldots] All the above vague statements have a single
goal---to point out although the simplex range
searching problem and related questions may look completely
solved, a really satisfactory solution may still await discovery.''
}

We actually find two different ways to achieve our improved result for Hopcroft's problem.
The first approach, based on a new form of fractional
cascading and described in Section~\ref{sec:cascade}, works only in 2D and is randomized, but can be adapted to yield efficient data structures 
for online halfplane range counting queries.  The second approach, based on
decision trees and described in Section~\ref{sec:dectree}, works in any constant dimension 
and is deterministic, but yields algorithms that are  far less practical (and, some might say, ``galactic''), although the ideas
are conceptually not complicated and are theoretically quite powerful.

\Paragraph{First approach via 2D fractional cascading.}
The extra factors in previous algorithms to Hopcroft's problem in 2D come from
the cost of multiple point location queries in arrangements of related subsets of lines.  We propose an approach based on
the well-known \emph{fractional cascading} technique of Chazelle and Guibas \cite{ChazelleG86,ChazelleG86ii}.  Fractional cascading
allows us to search for an element in multiple lists of elements in 1D, 
under certain conditions, in $O(1)$ time per list instead of $O(\log n)$.
The idea involves iteratively taking a fraction of the elements from one list and overlaying
it with another list.
Unfortunately, the technique does not extend to 2D point location in general:
overlaying two planar subdivisions of linear size may create a new planar
subdivision of quadratic size.
In fact, Chazelle and Liu~\cite{ChazelleL04} formally proved lower bounds in the pointer
machine model that rule out 2D fractional cascading even in very simple scenarios.
(Recently, Afshani and Cheng~\cite{AfshaniC20} obtained some new results
on 2D fractional cascading but only for \emph{orthogonal} subdivisions.)

We show that fractional cascading, under certain conditions, is still possible for 2D arrangements of lines!  The basic reason is that arrangements of lines already
have quadratic size to begin with, and overlaying two such arrangements still yields
an arrangement of quadratic size.  One technicality is that we now need randomization
when choosing a fraction of the lines.  We will incorporate standard techniques
by Clarkson and Shor~\cite{ClarksonS89} on geometric random sampling.

\Paragraph{Second approach via decision trees.}
Our second approach works very differently and proceeds by first bounding the algebraic 
decision tree complexity of the problem, i.e., we count only the cost of 
comparisons, and ignore all other costs.  Here, comparisons
refer to testing the signs of constant-degree polynomial functions on a constant
number of the input real numbers.  It was observed~\cite{Matousek93} that for
Hopcroft's problem, improved (nonuniform) decision tree bounds would automatically imply improved (uniform) time bounds,
because after
a constant number of levels of Matou\v sek's recursion~\cite{Matousek93}, the
input size can be made so tiny (e.g., $o(\log\log\log n)$) that we can afford
to precompute the entire decision tree for such tiny inputs. 
(In the algorithms literature, there are other examples of problems for which time complexity
has been shown to be equivalent to  decision tree complexity, such as matrix searching~\cite{Larmore90} and
minimum spanning trees~\cite{PettieR02}.  There are also examples of problems
for which polylogarithmic speedups of algorithms were obtained by first
considering the decision tree complexity, such as
all-pairs shortest paths~\cite{Fredman76SICOMP},
3SUM~\cite{GronlundP18}, and 
at least one other problem in computational geometry~\cite{Chan13}.)

In a seminal work, Fredman~\cite{Fredman76} showed that $M$
values can be sorted using just $O(M)$ comparisons instead of $O(M\log M)$, under certain scenarios when the values
``originate from'' a smaller set of numbers.  For example, the $x$-coordinates
of the $O(n^2)$ intersections of a set of $n$ lines can be sorted using
$O(n^2)$ comparisons~\cite{SteigerS95}, since these $O(n^2)$ values come from $O(n)$ real numbers.
(Another example from Fredman's original paper is the well-known \emph{$X+Y$ sorting problem}, although for the decision tree complexity of that particular problem,  simpler~\cite{SteigerS95} and better~\cite{KaneLM19} methods were later found.)
We show that this type of result is not limited to the sorting problem alone,
and that logarithmic-factor shavings in the decision tree setting are actually
not difficult to obtain for many other problems,
including point location of multiple query points in multiple subdivisions,
assuming that
the query points and subdivision vertices originate from $O(n)$ real numbers.
The improvement for Hopcroft's problem  then follows.
The technique works beyond 2D and is more general than fractional cascading
(which requires the subdivisions to be organized in the form of a bounded-degree
tree or dag).  In Appendix~\ref{app:dectree}, we mention applications of our framework to the decision tree complexities of other problems, although for these problems, we do not get improvements in time
complexities.

\Paragraph{Applications.}
In Sections~\ref{sec:var}--\ref{sec:app}, we describe some of the algorithmic consequences of our approaches, including:

\begin{itemize}
    \item An $O(n^{4/3})$-time algorithm for line segment intersection counting in 2D\@.  The best previous bound was $O(n^{4/3}\log^{1/3}n)$ by Chazelle~\cite{Chazelle93}.
    \item An $O(n^{4/3})$-time algorithm for computing the connected components among $n$ line segments in 2D\@.  The best previous bound was $O(n^{4/3}\log^3n)$ by Lopez and Thurimella~\cite{LopezT95} (a simpler alternative using biclique covers was noted by Chan~\cite{Chan06SICOMP} but still had extra logarithmic factors).
    \item An $O(n^{4/3})$-time randomized algorithm for bichromatic closest pair and Euclidean minimum spanning tree in 3D and in 4D\@.
    The best previous bound in 3D was
    $O(n^{4/3}\log^{4/3}n)$ (randomized) by Agarwal et al.~\cite{AgarwalES91}.
    \item An $O(n^{4/3})$-time algorithm for the \emph{line towering problem} in 3D (deciding whether some red line is below some blue line, given $n$ red lines and $n$ blue lines).     The best previous bound in 3D was
    $O(n^{4/3+\eps})$~\cite{ChazelleEGSS96}.
    \item An $O(n^{4/3})$-time randomized algorithm for the \emph{distance selection problem} in 2D (selecting the $k$-th distance among $n$ points).  The best previous randomized bound was $O(n^{4/3}\log^{2/3}n)$~\cite{MatIPL91} (see also
    \cite{AgarwalASS93,KatzS97,ChanIJCGA01,Wang22}).
\end{itemize}
The above list is not meant to be exhaustive; we expect more applications to follow.
The above improvements may appear a bit larger than for Hopcroft's problem, but to be fair, we should mention that existing techniques can already lower many of the extra factors from these previously stated bounds (though this may have been missed in previous papers).  Our new techniques are for removing the
remaining $2^{O(\log^*n)}$ factors.

Our approach via 2D fractional cascading, in combination with Fredman's decision tree technique, also leads to new randomized data structures for 2D halfplane range counting, for example, achieving
$P(n)=O(n^{4/3})$ expected preprocessing time and $Q(n)=O(n^{1/3})$ expected query time.
This removes a $2^{O(\log^*n)}$ factor from a previous result by Chan~\cite{Chan12}.   We can get  
the same bounds for ray shooting among $n$ line segments in 2D; see \cite{Chan12,Matousek93,Wang20} for previous work.
For this class of data structure problems in 2D, it is generally believed that $P(n)Q(n)^2=\Omega(n^2)$.  Known data structures achieve preprocessing/query trade-offs that almost
match the conjectured lower bound but with extra factors---our result is the first to eliminate all extra factors.

\section{Preliminaries}\label{sec:prelim}

We first review previous approaches to Hopcroft's problem. 
We concentrate on the version of the problem where we want to count the
number of point-above-hyperplane pairs, since counting incidence pairs can be easily reduced to this version (alternatively, our algorithms can be adapted directly to detect, count, or report incidence pairs).  For convenience,
we assume that the input is non-degenerate (admittedly, this assumption does not
make sense for the original problem about incidences, but 
it is straightforward to modify our algorithms to work in degenerate cases).

The main tool is the well-known Cutting Lemma:

\begin{lemma}\label{lem:cut}
\emph{(Cutting Lemma)}
Given $n$ hyperplanes in a constant dimension~$d$ and a parameter $r\le n$,
there exists a decomposition of $\R^d$ into $O(r^d)$ disjoint simplicial cells,
such that each cell is crossed\footnote{
Here, a hyperplane \emph{crosses} a cell if it intersects the interior of the cell.  If degeneracy is allowed, the statement of the lemma needs some modification (for example, allowing
non-full-dimensional simplicial cells, and defining ``crossing'' to mean
``intersecting but not containing'').
} 
by at most $n/r$ hyperplanes.  

The cells, the list of hyperplanes crossing each cell (i.e., the \emph{conflict list} of each cell),
and the number of hyperplanes completely below each cell
can be constructed in $O(nr^{d-1})$ time.
Furthermore, given a set of $m$ points, we can find the cell containing each point, in $O(m\log r)$ additional time.
\end{lemma}

The existence part was first proved by Chazelle and Friedman~\cite{ChazelleF90},
building on the pioneering work by Clarkson and Shor~\cite{Clarkson87,Clarkson88,ClarksonS89} on geometric random sampling.
The stated time bounds were achieved by a construction of Matou\v sek~\cite{Matousek95}
for the case $r\le n^{1-\eps}$, and subsequently by
a construction of Chazelle~\cite{Chazelle93} for all $r\le n$.
The constructions are simpler if randomization is allowed.
(See also~\cite{ChanT16} for another deterministic construction in the 2D case.)

Chazelle's construction~\cite{Chazelle93} actually builds a hierarchy of cuttings,
yielding a strengthened lemma that is useful in some applications:

\begin{lemma}\label{lem:cut:hier}
\emph{(Hierarchical Cutting Lemma)}
Given $n$ hyperplanes in a constant dimension~$d$ and a parameter $r\le n$,
for some constant $\rho>1$,
there exists a sequence $\Xi_1, \ldots, \Xi_k$ with $k={\up{\log_\rho r}}$, where each $\Xi_i$ is a decomposition of $\R^d$ into $O((\rho^i)^d)$ disjoint simplicial cells each crossing at most
$n/\rho^i$ hyperplanes, and each cell in $\Xi_i$ is a union of $O(1)$
disjoint cells in $\Xi_{i+1}$.  All the cells can be constructed
in $O(nr^{d-1})$ time.
\end{lemma}

Given $m$ points and $n$ hyperplanes in  
a constant dimension~$d$,
let $T(m,n)$ be the time complexity of the problem of counting the number of point-above-hyperplane pairs.  We can solve the problem
by applying the Cutting Lemma:
Subdivide the cells so that each cell
contains at most $m/r^d$ points.
Since $O(r^d)$ extra vertical cuts in total suffice,
the number of cells remains $O(r^d)$.
For each cell $\D$, solve the subproblem 
for all points in $\D$ and all hyperplanes crossing $\D$.
Add to the counter the product of the number of points in $\D$ with the number of
hyperplanes below $\D$.
It follows that
\begin{equation}\label{eqn:recur}
 T(m,n) \:=\: O(r^d)\, T(m/r^d, n/r) + O(nr^{d-1} + m\log r).
\end{equation}


In the asymmetric setting when $m$ is much larger than $n$,
we have $T(m,n)=O(n^d + m\log n)$, for example, by setting $r=n$ above;
in the 2D case, this time bound can be obtained more directly by 
answering $m$ planar point location queries~\cite{BergCKO08,Kirkpatrick83} in the arrangement of $n$ lines,
which has $O(n^2)$ size and can be constructed in $O(n^2)$ time~\cite{BergCKO08,EdelsbrunnerOS86} (since we can label each face of the arrangement with the number of lines below it).
By point-hyperplane duality, we also have $T(m,n)=O(m^d + n\log n)$.
Putting this bound into (\ref{eqn:recur}) gives
\begin{equation}\label{eqn:sol1}
 T(m,n) \:=\: O(r^d)\, [(m/r^d)^d + (n/r)\log n] + O(nr^{d-1}+m\log r).
\end{equation}
Setting $r=n^{1/(d+1)}$ then yields $T(n,n) = O(n^{2d/(d+1)}\log n)$ in the
symmetric case $m=n$.

Slight modification of the choice of $r$ can lower the $\log n$ factor a bit
(as was done by Chazelle~\cite{Chazelle93}),
but Matou\v sek~\cite{Matousek93} proposed a better algorithm using $O(\log^*n)$
levels of recursion.  The following is a slightly cleaner rederivation of Matou\v sek's recursion:

First, by point-hyperplane duality, we have the following recurrence:
\begin{equation}\label{eqn:recur2}
 T(m,n) \:=\: O(r^d)\, T(m/r, n/r^d) + O(mr^{d-1} + n\log r).
\end{equation}
By applying (\ref{eqn:recur}) and (\ref{eqn:recur2}) in succession and letting $T(n)=T(n,n)$,
\[ 
 T(n)\:=\: O(r^{2d})\, T(n/r^{d+1}) + O(nr^{d-1}\log r),
\] 
assuming that $r\le n/r^d$.
By choosing $r=(n/\log^C n)^{1/(d+1)}$ for a sufficiently large constant~$C$, 
\begin{equation}\label{eqn:logstar}
 T(n) \:=\: O(n/\log^C n)^{2d/(d+1)} T(\log^C n) + o(n^{2d/(d+1)}).
\end{equation}
This recurrence then solves to $T(n)=T(n,n)\le n^{2d/(d+1)}2^{O(\log^*n)}$.

Improving the remaining iterated logarithmic factor requires
new ideas.  In the next two sections, we propose two different approaches to do so, the first of which works in the 2D case.

\section{First Approach via 2D Fractional Cascading}\label{sec:cascade}

For a set $L$ of lines, let $\A(L)$ denote the arrangement of $L$.
Let $\VD(L)$ denote the \emph{vertical decomposition} of $\A(L)$, 
where the faces of $\A(L)$ are divided into trapezoids by drawing vertical
line segments at the vertices.

\subsection{Fractional cascading for arrangements of lines}

We begin by introducing a subproblem about point location in multiple
arrangements of lines:

\begin{problem}\label{prob:cascade}
Let $T$ be a rooted tree with maximum degree $c_0=O(1)$. 
Each node $u\in T$ stores a set $L_u$ of at most $z$ lines in $\R^2$.
We want to answer the following query: for a given query point $q\in\R^2$
and a subtree $T_q$ of $T$ containing the root, output for each node $u\in T_q$ (a label of) the face of $\A(L_u)$ containing $q$.
\end{problem}

In the 1D case where each set $L_u$ consists of points on the real line, the fractional cascading technique~\cite{ChazelleG86,ChazelleG86ii} gives a solution
with $O(\log z + |T_q|)$ query time, after $O(|T|z\log(|T|z))$ preprocessing time
(or actually $O(|T|z)$ if each set $L_u$ has been pre-sorted).
However, in 2D, generalizations of fractional cascading for
non-orthogonal problems were not
known before.  Answering each of the $|T_q|$ point location queries~\cite{BergCKO08} 
separately would give $O(|T_q|\log z)$ query time and $O(|T|z^2)$ preprocessing time.  

By combining the original fractional cascading technique with some new
simple ideas, we show how to improve the query time
to $O(\log z + |T_q|)$ for the 2D problem, matching the 1D bound, and effectively achieving $O(1)$ cost per point location query, after
an initial $O(\log N)$ cost.  One caveat is that we require (Las Vegas)
randomization, and assume the query points and subtrees are oblivious to the
random choices made by the data structure.
A more crucial caveat is that we assume the queries are
\emph{$x$-monotone}, i.e., the query points arrive in
increasing order of $x$-coordinates.  In the setting of offline queries 
when all the query points are given in advance (as is the case in our
application to Hopcroft's problem later), this assumption can be satisfied
by pre-sorting the query points and processing the queries in that order.

\begin{lemma} \label{lem:cascade}
For the case of $x$-monotone queries,
there is a randomized data structure for Problem~\ref{prob:cascade}
with $O(|T|z^2)$ amortized expected preprocessing time, such that a query
for a fixed point $q$ and subtree $T_q$ takes 
$O(\log z+ |T_q|)$ amortized expected time.
\end{lemma}

\begin{myproof}
  For each node $u\in V$, we construct a set of lines $L_u^+$ starting at the leaves and proceeding bottom-up:
  If $u$ is a leaf, let $L_u^+ = L_u$. 
  If $u$ is an internal node, then for each child $v$ of $u$, sample a random subset $R_v\subset L_v^+$ of size $|L_v^+|/c$ for some constant~$c$. 
  Let $L_u^+= L_u \,\cup\, \bigcup_{\mbox{\scriptsize child $v$ of $u$}}R_v$.

  Letting $s_i$ denote the maximum size of $L_u^+$ over all nodes $u$ at level $i$, 
  we obtain the recurrence $s_i \le z + c_0s_{i+1}/c$, which yields $s_i=O(z)$
  by choosing a constant $c>c_0$.  Thus, $|L_u^+|=O(z)$ at every node $u$.


During preprocessing, we do the following:
\begin{enumerate}
\item For every node $v$, we construct the arrangements $\A(L_v)$,
$\A(L_v^+)$, and $\A(R_v)$, and the vertical decomposition $\VD(R_v)$, in $O(|L_v^+|^2)=O(z^2)$ time.
\item For each node $v$ and each trapezoid $\D\in \VD(R_v)$, we store the
\emph{conflict list} of $\D$, i.e., the subset
$(L_v^+)_\D$ of all lines in $L_v^+$ crossing $\D$, and we construct
the sub-arrangement $\A((L_v^+)_\D)$ inside $\D$, and link its features 
to those of the overall arrangement $\A(L_v^+)$.
\item For each node $u$, since $L_u\subset L_u^+$, 
each face of $\A(L_u^+)$ is contained in a unique face of $\A(L_u)$;
we store a pointer linking the former to the latter.
Similarly, for each child $v$ of $u$, since $R_v\subset L_u^+$,
each face of $\A(L_u^+)$ is contained in a unique face of $\A(R_v)$;
we store a pointer linking the former to the latter.
\item At the root $w$, we store $\A(L_w^+)$ in a known point location
data structure~\cite{BergCKO08,Kirkpatrick83}.
\end{enumerate}
By a standard analysis of Clarkson and Shor~\cite{ClarksonS89}, $\E[\sum_{\Delta\in \VD(R_v)} |(L_v^+)_\Delta|^2] = O(|L_v^+|^2)=O(z^2)$, and 
step~2 can be done in $O(z^2)$
expected time.  Step~3 can also be done in $O(z^2)$ time by traversing
the faces in these arrangements.
Thus, the total expected preprocessing time is $O(|T|z^2)$.

To answer a query for a point $q$ and subtree $T_q$, 
we first find the face of $\A(L_w^+)$ containing $q$ at the root $w$
in $O(\log z)$
time by the point location structure from step~4.
We consider each node $u\in T_q$ starting at the root and proceeding 
top-down (for example, in depth-first or breadth-first order).
Suppose we have already found the face $f$ of $\A(L_u^+)$ containing $q$.
By following one of the pointers from step~3, we know the face of $\A(L_u)$ containing $q$ and can output the result.
Next, take each child $v$ of $u$ in $T_q$.
\begin{enumerate}
\item[(i)] By following one of the pointers from step~3, we also know the face $f'$ of $\A(R_v)$ containing $q$.
\item[(ii)] Find the trapezoid $\D$ in $\VD(R_v)$ containing $q$.  This can be done by searching in the vertical decomposition of the face $f'$.
\item[(iii)] Find the face $f''$ of $\A(L_v^+)$ containing $q$.  This can be done by searching in the sub-arrangement $\A((L_v^+)_\D)$ inside $\D$.
\end{enumerate}
We can now repeat the process at the child node~$v$.  See Figure~\ref{fig_frac_casc}.

\begin{figure}
    \centering
    \includegraphics[scale=0.78]{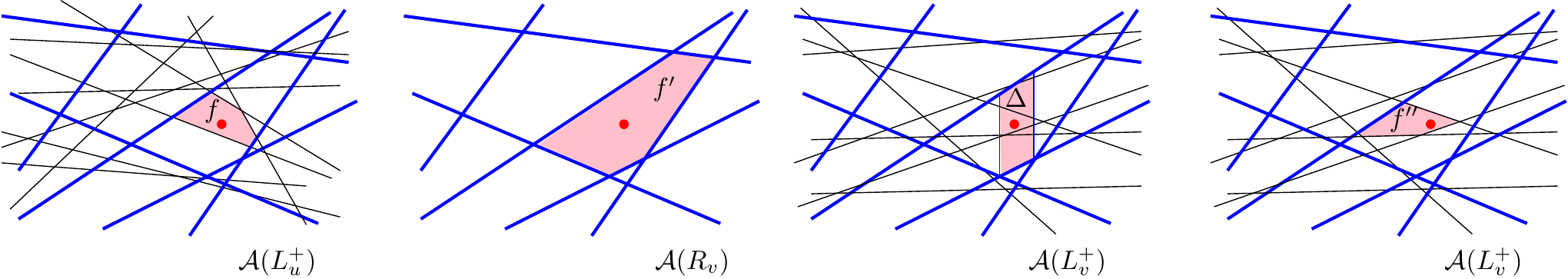}
    \caption{Fractional cascading for 2D arrangements of lines.}
    \label{fig_frac_casc}
\end{figure}

Step~(iii) can be done in $O(|(L_v^+)_\D|)$ time by naive linear searches
in the conflict list $(L_v^+)_\D$.
By a standard analysis of Clarkson and Shor~\cite{ClarksonS89} (e.g., 
see \cite[proof of Lemma~6]{ChanHN20arxiv}),
$\E[|(L_v^+)_\D|]=O(c)$ for the cell $\D\in \VD(R_v)$ containing $q$
(assuming that $q$ is independent of $R_v$).  Thus, step~(iii) takes constant expected time.

Step~(ii) may require nonconstant time, since the face $f'_v$ (a convex polygon) may
have large complexity.  Here, we will exploit the $x$-monotone query assumption.  For each face $f'$ of $\VD(R_v)$,
we maintain a pointer $p_{f'}$ in the $x$-sorted list of the vertices of $f'$.  Initially, the pointer is at the leftmost vertex of $f'$.  To perform step~(ii), we simply do a linear search in the list, starting from the previous position of
$p_{f'}$ and advancing  $p_{f'}$ from left to right,
till we encounter a vertex to the right of~$q$.  The total time of
all linear searches in $f'$ is proportional to the complexity of $f'$.
Summing over all faces $f'$ of $\VD(R_v)$ gives $O(|R_v|^2)=O(z^2)$, and so
the cost is absorbed by the preprocessing cost by amortization.

We conclude that the amortized expected cost of a query is $O(\log z+|T_q|)$.
\end{myproof}

\Paragraph{Remarks.}
Randomized variants of 1D fractional cascading  have been considered before
(e.g., see \cite{ChanN18}), but the idea of
combining fractional cascading with
Clarkson--Shor-style random sampling appears new.

The approach can be generalized to dags with bounded out-degree, and
to unbalanced settings where the given sets $L_u$ may have different sizes,
but such generalizations will not be needed in our application to Hopcroft's problem.

The approach doesn't seem to work for point location in other
geometric structures (such as planar Voronoi diagrams), since the face containment property in step~3 holds only for arrangements.
And the approach doesn't seem to work for higher-dimensional hyperplane
arrangements, since the assumption of $x$-monotone queries alone doesn't seem 
to help in step~(ii).  (So, we are fortunate that the approach works at all for
2D line arrangements!)

\subsection{Hopcroft's Problem in 2D}

We now use our new 2D fractional cascading technique to solve
Hopcroft's problem:

\begin{theorem}\label{thm:hopcroft:2d}
Given $n$ points and $n$ lines in $\R^2$, we can count the number of point-above-line
pairs in $O(n^{4/3})$ expected time.
\end{theorem}
\begin{myproof}
Recall that the Cutting Lemma
gives $O(r^2)$ subproblems, each with $O(n/r^2)$ points and $O(n/r)$
lines.  When we derived (\ref{eqn:sol1}) in Section~\ref{sec:prelim}, 
we solve each subproblem directly by building the arrangement
of the $O(n/r^2)$ dual lines and answering a point location query for each of the $O(n/r)$ dual points.  We will now use Lemma~\ref{lem:cascade}
to speed up point location.  In particular, recursion is not needed!

More precisely, apply the Hierarchical Cutting Lemma, which in $O(nr)$ time
generates a tree of cells with degree $O(1)$ and height $k=\up{\log_\rho r}$
(where $\Xi_i$ corresponds to the cells at level~$i$ of the tree).
Subdivide the leaf cells, in the form of a binary tree,
so that each leaf cell has at most $n/r^2$ points.
Since $O(r^2)$ vertical cuts in total suffice,
the number of nodes remains $O(r^2)$, and the maximum degree of the tree remains $O(1)$.

For each node $v$, let $P_v$ be the set of all input points inside $v$'s cell,
and $L_v$ be the set of all input lines crossing $v$'s cell.
Let $P_v^*$ be the dual of $P_v$ (a set of lines)
and $L_v^*$ be the dual of $L_v$ (a set of points).
The problem is solved by finding the face of $\A(P_v^*)$
containing each point in $L_v^*$, for each leaf node $v$.
This is an instance of Problem~\ref{prob:cascade}
for a tree $T$ with $|T|=O(r^2)$ nodes,
where the set of lines at each leaf node~$v$ is $P_v^*$, which has size at most
$z := n/r^2$,
and the set of lines at each internal node is the empty set.
The query points are dual of the input lines.  The subtree $T_{\ell^*}$ corresponding to the query point $\ell^*$ consists of all nodes
whose cells are crossed by the input line~$\ell$---these nodes indeed form
a subtree containing the root, since if a cell is crossed by $\ell$,
all ancestor cells are crossed by $\ell$.  As mentioned,
in the offline setting, we can process the query points in increasing $x$-order to satisfy the assumption of
$x$-monotone queries.

Recall that for each $i=1,\ldots,k$,
there are $O(\rho^{2i})$ cells at level $i$, each crossed by $O(n/\rho^i)$ lines; and there are $O(r^2)$ cells at levels beyond $k$, each crossed
by $O(n/r)$ lines.
It follows that $\sum_\ell |T_{\ell^*}|=O(\sum_{i=1}^k \rho^{2i}\cdot n/\rho^i + r^2\cdot n/r) = O(nr)$.
By Lemma~\ref{lem:cascade}, the total expected cost of all
$n$ queries is 
\[ O\left(|T|z^2 + n\log z + \sum_\ell |T_{\ell^*}|\right)
\:=\: O(r^2 (n/r^2)^2 + n\log n + nr).
\]
Setting $r=n^{1/3}$ yields $O(n^{4/3})$.
\end{myproof}

\IGNORE{
\begin{corollary}\label{cor:hopcroft:2d}
Given $m$ points and $n$ lines in 
a constant dimension $d\ge 2$, we can count the number of point-above-hyperplane
pairs in $O((mn)^{2/3}+m\log n + n\log m)$ time.
\end{corollary}
\begin{myproof}
Suppose $m\ge n$ (if not, we can apply point-line duality).  
In (\ref{eqn:recur}),
choose $r=m/n$ and put $T(n^2/m,n^2/m)=O((n^2/m)^{4/3})$.
Then $T(m,n)=O((mn)^{2/3} + m\log n)$.
This assumes $m\le n^2$,
but for $m>n^2$, we can switch to
a known $O(n^2 + m\log n)$-time algorithm.
\end{myproof}
}

\Paragraph{Remark.}
It is possible to derandomize the algorithm (by first making the problem size tiny by using a constant number of rounds of Matou\v sek's recursion~\cite{Matousek93}, and doing brute-force search to find a sequence of deterministic choices that work for all inputs).  
We will omit the details, since the second approach in the next section
is automatically deterministic, and a complicated derandomization of
the first approach would defeat its main virtue, simplicity.

\section{Second Approach via Decision Trees}\label{sec:dectree}

In this section, we propose a different approach to solve Hopcroft's problem,
which works in any constant dimension.  The approach is based on
a general framework for bounding decision tree complexities.

\subsection{Framework}\label{sec:framework}

\newcommand{\xx}{\boldsymbol{x}}
For most comparison-based algorithms in computational geometry,
the input can be described by a vector of $N$ real numbers $\xx=(x_1,\ldots,x_N)$, and the only primitive operations needed on the input real numbers are tests of the form, ``is $\gamma(\xx)$ true?'', for a predicate $\gamma$
from some collection $\Gamma$.
We call such a test a \emph{$\Gamma$-comparison}, and such an algorithm a \emph{$\Gamma$-algorithm}.  
To study the decision tree complexity of problems,
we are interested in bounding the number of $\Gamma$-comparisons made by a $\Gamma$-algorithm, ignoring all other costs.  A predicate $\gamma$ can be equivalently viewed as a set in $\R^N$,
namely, the set of all inputs $\xx$ such that $\gamma(\xx)$ is true. We make the following assumptions:
\begin{itemize}
    \item  each $\gamma\in\Gamma$ is a semialgebraic set of constant degree and constant complexity, i.e., we can test whether $\gamma(\xx)$ is true by evaluating the signs of $O(1)$ polynomials of degree $O(1)$ in  $O(1)$ number of the variables $x_1,\ldots,x_N$; 
    \item the number of possible  comparisons is polynomially bounded, i.e., $|\Gamma|\le N^{O(1)}$.  
\end{itemize}
We say that $\Gamma$ is \emph{reasonable} if the above conditions are satisfied.
(For example, many algorithms for 2D convex hulls use only orientation tests for triples of input points, and thus fit the above framework, using a reasonable collection $\Gamma$ of $O(N^3)$ semialgebraic sets of degree~2.)

Consider the arrangement $\A(\Gamma)$ of the semialgebraic sets in $\Gamma$, living in a high-dimensional space $\R^N$.
Each cell of $\A(\Gamma)$ can be associated with a sign vector, where the $i$-th component is $+1$ if points in the cell satisfy the $i$-th predicate of $\Gamma$, and $-1$ otherwise.
Points in the same cell correspond to inputs that have the same outcomes with respect to all possible comparisons.  (For example, if $\Gamma$ corresponds to orientation tests in 2D, then cells correspond to the standard notion of \emph{order types}~\cite{GoodmanP86}.)
Naively, the number of different signed vectors is bounded by  $2^{|\Gamma|} \le 2^{N^{O(1)}}$.  However, the Milnor--Thom Theorem \cite{Milnor64,Thom65,AgarwalS00} provides a better bound: the number of cells of $\A(\Gamma)$, and thus the number of signed vectors, are in fact at most
$O(|\Gamma|)^N \le N^{O(N)}$.

Suppose that an algorithm has made some number of comparisons, say, defined by  $\gamma_1,\ldots,\gamma_\ell\in \Gamma$ with outcomes $b_1,\ldots,b_\ell$.
We call a cell of $\A(\Gamma)$ \emph{active} if it is consistent with the outcomes of the comparisons made so far, i.e., for all $\xx$ in the cell, $\gamma_i(\xx) = b_i$  for all $i=1,\ldots,\ell$.

\newcommand{\Act}{\Pi_{\mbox{\scriptsize\rm active}}}

Let $\Act$ denote the set of all active cells.
Define the \emph{potential} \[\Phi=\log|\Act|.\] 
(Intuitively, one can view this as an information-theoretic lower bound on the  
depth of an algebraic decision tree needed to determine which cell of $\Act$ the input is in.)
At the beginning, all cells of $\A(\Gamma)$ are active, and $|\Act|\le N^{O(N)}$, and so $\Phi=O(N\log N)$.
As the algorithm progresses and makes more comparisons, the number of active cells can only decrease, and so $\Phi$ can only decrease.
For any operation or subroutine, we use the notation $\Delta\Phi$ to denote
the change in potential, i.e., the value of $\Phi$ after the operation minus the value of $\Phi$ before the operation.  Since $\Phi$ can only decrease, $\Delta\Phi$ is always negative or zero.  

Inspired by \emph{amortized analysis}, we design $\Gamma$-algorithms by adopting the following philosophy: if an operation is costly but allows us to shrink the number of active cells and thus decrease $\Phi$, the work done may still be worthwhile
if we can charge the cost to the potential decrease $-\Delta\Phi$.  Individual $-\Delta\Phi$ terms are easy to add up, by a telescoping sum, and the total is equal to the global $-\Delta\Phi$, which is bounded by $O(N\log N)$.


The following simple but crucial lemma provides the basic building block behind our algorithms:



\begin{lemma}\label{lem:choice}
\emph{(Basic Search Lemma)}
Consider the $\Gamma$-algorithm framework defined above.
Let $\gamma_1,\ldots,\gamma_r\in\Gamma$.  Suppose we are promised that
$\gamma_1(\xx)\vee\cdots\vee \gamma_r(\xx)$ is true for all input $\xx$ in the active cells.
Then we can search for a $k\in\{1,\ldots,r\}$ such that $\gamma_k(\xx)$ is true, by making $O(1-r\Delta\Phi)$ $\Gamma$-comparisons.
\end{lemma}
\begin{myproof}

Pick an index $k$ such that the number of active cells satisfying $\gamma_k$ is
at least $|\Act|/r$ (we know that $k$ exists because of the promise).
Make the comparison~$\gamma_k$.
If $\gamma_k(\xx)$ is true, we are done.  If $\gamma_k(\xx)$ is false, cells that satisfy $\gamma_k$ are no longer active, and so the number of active cells
after the comparison is at most $\frac{r-1}r |\Act|$.  The potential $\Phi$ thus decreases by at least $\log\frac{r}{r-1}=\Omega(\frac 1r)$.
Now repeat.  The number of iterations is bounded by $O(-r\Delta\Phi)$.  
%
\end{myproof}

The above lemma is useful as it provides a mild form of nondeterminism, allowing us to ``guess''
which one of $r$ choices is correct, with just $O(1)$ \emph{amortized} cost instead of $O(r)$.
This scenario arises naturally in point location:
searching for which of $r$ cells contains a given point.

We emphasize that the above lemma does not guarantee good running time.
As preprocessing, we can first construct the arrangement $\A(\Gamma)$~\cite{AgarwalS00,BasuPR96STOC,BasuPR96}.  By scanning all its cells and their sign vectors, we can determine which cells are active, and which cell satisfies $\gamma_k$.  This process is time-consuming, since the size of $\A(\Gamma)$ and the size of $\Act$ are huge: the construction time for the $N$-dimensional arrangement $\A(\Gamma)$ is 
$2^{N^{O(1)}}$, and the time to scan all the cells is
$O(|\Act|)= N^{O(N)}$.
But such computation does not require comparisons on the input $\xx$, and has zero cost in the decision tree setting.
The manipulation of such large sets $\Pi$ is precisely what makes the framework different from traditional algorithms.

\subsection{Example: Fredman's sorting result}\label{sec:sort}

In his seminal paper, Fredman~\cite{Fredman76} showed surprisingly that $M$ values can be sorted
using $O(M)$ comparisons instead $O(M\log M)$ under certain scenarios, when the $M$ values ``originate
from'' a smaller set of $N$ input numbers.  To provide a warm-up example illustrating the usefulness of
the Basic Search Lemma (Lemma~\ref{lem:choice}), we rederive (a slightly weaker version of) Fredman's result by a quick simple proof.

\begin{theorem}\label{thm:sort}
We can sort $M$ values $p_1,\ldots,p_M$
using $O(M-M^\eps\Delta\Phi)\le O(M+M^\eps N\log N)$
$\Gamma$-comparisons  for any constant $\eps>0$, assuming that testing whether $p_i< p_j\le p_k$ can be expressed as a $\Gamma$-comparison
on the input $\xx\in\R^N$, and assuming that $\Gamma$ is reasonable.
\end{theorem}
\begin{myproof}
We sort by performing $M$ repeated insertions.
In the $k$-th insertion, we need to find the predecessor of $p_k$ among the sorted re-ordering of $p_1,\ldots,p_{k-1}$.  
We can find the predecessor of $p_k$ among $r$ quantiles using $O(1-r\Delta\Phi)$
$\Gamma$-comparisons by an immediate application of the Basic Search Lemma.
Therefore, with $O(\log_r M)$ levels of recursion, we can find the predecessor of $p_k$ in
a sorted list of size $O(M)$ using $O(\log_r M - r\Delta\Phi)$ $\Gamma$-comparisons.

The total number of comparisons is thus $O(M\log_r M - r\Delta\Phi)$.
Choose $r=M^\eps$.  
\end{myproof}

As one application of the theorem, we immediately obtain the following:

\begin{corollary}
We can sort the $x$-coordinates of all $O(n^2)$ vertices
of an arrangement of $n$ given lines in the plane by an algebraic decision tree 
of $O(n^2)$ depth.
\end{corollary}

\begin{myproof}
The $M=O(n^2)$ vertices of the arrangement are intersections of pairs of lines, which are defined by $N=2n$ slope/intercept values. Testing if the $x$-coordinate of a vertex is between two other vertices can be done using a reasonable collection $\Gamma$ of $O(N^4)$ predicates of constant degree (where the variables are the slopes/intercepts of the input lines). The conclusion follows from Theorem~\ref{thm:sort}.
\end{myproof}

\Paragraph{Remark.} Fredman's sorting bound actually eliminated the extra $M^\eps$ factor in Theorem~\ref{thm:sort}.  This is achieved by finding predecessors
using a weighted binary search instead of Lemma~\ref{lem:choice}; see Appendix~\ref{app:dectree}.  In our main applications, the $\Delta\Phi$ terms will not
dominate, so there is no need to optimize the $M^\eps$ factor.

\subsection{Hopcroft's problem in any constant dimension}\label{sec:hopcroft:dectree}

We now apply our framework to bound the decision tree complexity of Hopcroft's problem in any constant dimension~$d$.

We first solve the asymmetric case of the problem when the number $m$ of points  is much
larger than the number  $n$ of hyperplanes.  As noted in Section~\ref{sec:prelim}, a known solution has
$O(n^d + m\log n)$ complexity.  We eliminate the logarithmic factor and get
$O(n^d + m)$ amortized cost:

\begin{lemma}\label{lem:hopcroft:dectree}
Given a set $P$ of $m$ points and a set $H$ of $n$ hyperplanes in 
a constant dimension~$d$, we can count the number of point-above-hyperplane
pairs 
using $O(n^d + m - n^\eps\Delta\Phi)$ $\Gamma$-comparisons for any constant $\eps>0$,
assuming that certain primitive operations on the points and hyperplanes
can be expressed as $\Gamma$-comparisons.
\end{lemma}
\begin{myproof}
We apply the Cutting Lemma 
in the same way that we derived (\ref{eqn:recur}) in Section~\ref{sec:prelim}, 
but with one change:
To find which one of the $O(r^d)$ cells contains each given point, 
we apply the Basic Search Lemma, which makes $O(1-r^d\Delta\Phi)$ $\Gamma$-comparisons per point (instead of $O(\log r)$).
This assumes that deciding whether a point of $P$ lies in one of the cells can be expressed as a $\Gamma$-comparison.
Excluding the $O(-r^d\Delta\Phi)$ terms, the number of comparisons made satisfies the following recurrence: 
\begin{equation}\label{eqn:recur:dectree}
 T'(m,n) \:=\: O(r^d)\, T'(m/r^d, n/r) + O(nr^{d-1} + m).
\end{equation}

We use a fixed value of $r$ for the whole recursion, assuming that $n$ is a power of $r$.  In the base case $n=r$, we get $T'(m,r)=O(r^d+m)$ 
without needing the Cutting Lemma (instead, just using a triangulation of the arrangement $\A(H)$, which can be constructed in $O(r^d)$ time~\cite{EdelsbrunnerOS86,EdelsbrunnerSS93}).

The recurrence solves to $T'(m,n)=O((n^d + m) \cdot 2^{O(\log_r n)})$.
Choosing $r=n^{\eps/d}$ (with $d/\eps$ an integer) gives 
a bound of $O(n^d + m - n^\eps\Delta\Phi)$ on the total number of comparisons made.

We note that since $n$ is a power of $r$, the applications
of the Cutting Lemma here have $r\le\sqrt{n}$,
and so Matou\v sek's cutting construction~\cite{Matousek95} suffices,
which can be implemented with predicates of constant complexity.
\end{myproof}

The above improvement leads to improvement in the symmetric case:

\begin{lemma}\label{lem:hopcroft:dectree2}
Given $n$ points and $n$ hyperplanes in 
a constant dimension $d\ge 2$, we can count the number of
point-above-hyperplane pairs by an algebraic decision tree of $O(n^{2d/(d+1)})$ depth.
\end{lemma}
\begin{myproof}
By point-hyperplane duality,  Lemma~\ref{lem:hopcroft:dectree}
implies $T'(m,n)=O(m^d+n)$, excluding $O(-m^\eps \Delta\Phi)$ terms.
We obtain the following consequence
of (\ref{eqn:recur:dectree}): 
\[ T'(m,n) \:=\: O(r^d)[(m/r^d)^d + n/r] + O(nr^{d-1} + m).
\]
Setting $m=n$ and $r=n^{1/(d+1)}$ gives
$T'(n,n) = O(n^{2d/(d+1)})$.
The excluded terms sum to $O(-n^\eps \Delta\Phi)\le O(n^\eps\cdot N\log N)$,
which does not dominate (as $N=O(n)$).  

Since $r\ll n^{1-\eps}$, Matou\v sek's cutting construction~\cite{Matousek95}
suffices, and the whole algorithm can be implemented with a reasonable collection $\Gamma$ of predicates.
\end{myproof}

For Hopcroft's problem, an improvement in the
decision tree complexity can be converted to an improvement in
time complexity, 
as already pointed out in Matou\v sek's paper~\cite{Matousek93}
(Matou\v sek acknowledged David Eppstein for the idea, and this type
of argument has appeared before in other contexts, e.g.,~\cite{Larmore90}).  We redescribe the argument below: 

\begin{theorem}\label{thm:hopcroft}
Given  $n$ points and $n$ hyperplanes in 
a constant dimension $d\ge 2$, we can count the number of point-above-hyperplane
pairs in $O(n^{2d/(d+1)})$ time.
\end{theorem}
\begin{myproof}
Applying (\ref{eqn:logstar})
(from Section~\ref{sec:prelim}) three times gives
\begin{equation}\label{eqn:final}
 T(n) = O(n/b)^{2d/(d+1)}T(b) + o(n^{2d/(d+1)})
\end{equation}
with $b=O((\log\log\log n)^C)$.

Since the new input size $b$ is tiny, we can afford to build
the decision tree from Lemma~\ref{lem:hopcroft:dectree2} in advance, and
get $T(b)=O(b^{2d/(d+1)})$. 

Viewed another way: we can directly simulate the algorithm from Lemma~\ref{lem:hopcroft:dectree2} when the input size $b$ is tiny.  The 
the active cells $\Act$ have size
$O(b^c)^{O(b)}$ and can be initially computed in $2^{b^{O(1)}}$ time, as noted in Section~\ref{sec:framework}.
 For $b=O((\log\log\log n)^C)$,
these bounds are 
 smaller than $\log^{o(1)}n$, and in particular, we can encode and pack $\Pi$ in a single word with
$\log^{o(1)}n$ bits.  Each operation on $\Act$ used in the proof of
the Basic Search 
Lemma can be carried out in constant time by table lookup,
after a one-time preprocessing in $2^{\log^{o(1)}n}$ time.  This justifies $T(b)=O(b^{2d/(d+1)})$.

As a result, (\ref{eqn:final}) implies $T(n)= O(n^{2d/(d+1)})$.
\end{myproof}

\Paragraph{Remark.}
In the application of the Cutting Lemma in the above theorem, we now need
the case when $r$ is close to $n$
(when we apply (\ref{eqn:recur2}), $n$ becomes $n/r^d$,
and $r$ is close to $n/r^d$ for our choice of $r$ near $n^{1/(d+1)}$, 
ignoring polylogarithmic factors).  We can use
Chazelle's hierarchical cutting construction here~\cite{Chazelle93}.
However, 
one disadvantage is that  it generates cells whose
coordinates may have non-constant degree, 
for\footnote{
For $d=2$, this is not an issue if we use vertical decompositions in place
of bottom-vertex triangulations in Chazelle's construction.  All cells
would then be trapezoids, where the top or bottom side is defined by an input line and
the left or right vertical side has $x$-coordinate defined by a pair of
input lines.
} 
$d\ge 3$,
 though technically this is allowed in the real RAM model of computation.  


\section{Variants}\label{sec:var}

\subsection{Asymmetric case}

The asymmetric case reduces back to the symmetric case:

\begin{corollary}\label{cor:hopcroft}
Given $m$ points and $n$ hyperplanes in 
a constant dimension $d\ge 2$, we can count the number of point-above-hyperplane
pairs in $O((mn)^{d/(d+1)}+m\log n + n\log m)$ time.
\end{corollary}
\begin{myproof}
Suppose $m\ge n$ (if not, apply point-hyperplane duality).  
In (\ref{eqn:recur}),
choose $r$ so that $m/r^d=n/r$, i.e., $r=(m/n)^{1/(d-1)}$,
and put $T(m/r^d,n/r)=O(((m/r^d)(n/r))^{d/(d+1)})$.
Then $T(m,n)=O((mn)^{d/(d+1)} + m\log n)$.
This assumes $m\le n^d$,
but for $m>n^d$, we can switch to
a known $O(n^d + m\log n)$-time algorithm.
\end{myproof}

\subsection{Individual counts and sums of weights}

Our algorithms can be modified to output individual counts per point and hyperplane (i.e., the number of hyperplanes below each point, and the number of points above each hyperplane); consequently, our algorithms
can be used to
answer an offline sequence of $m$ halfspace range counting queries.
We can also generalize counting to summing weights.

\begin{lemma}\label{lem:arrang}
Given an arrangement of $n$ hyperplanes in $\R^d$ where each hyperplane has a weight and  each face has a weight, we can compute (a)~the sum of the weights of the hyperplanes below each face, and (b)~the sum of the weights of the faces above each hyperplane, in $O(n^d)$ total time.
\end{lemma}
\begin{myproof}
(a) is straightforward by traversing the faces of the arrangement.
For (b), we may assume that the input weights are at vertices instead of faces, by mapping each face to its highest vertex.
For each vertex, assign it to the line through one of its incident edges.
For each of the $O(n^{d-1})$ lines $\ell$, compute all prefix/suffix sums of the weights of the vertices assigned to $\ell$; the total time is $O(n^{d-1}\cdot n)$.  For each hyperplane $h$, we can then compute its answer by inspecting $O(n^{d-1})$ prefix/suffix sums; the total time is $O(n\cdot n^{d-1})$.
\end{myproof}

\begin{theorem}\label{thm:hopcroft:individual}
Given  $m$ weighted points and $n$ weighted hyperplanes in 
a constant dimension $d\ge 2$, we can sum the weights of hyperplanes
below each point, and the weights of points above each hyperplane,
in $O((mn)^{d/(d+1)}+m\log n + n\log m)$ total time.
\end{theorem}
\begin{myproof}
We modify the algorithm in Section~\ref{sec:hopcroft:dectree} (the 2D algorithm in Section~\ref{sec:cascade} can also be modified):

In the derivation of (\ref{eqn:recur}) and (\ref{eqn:recur:dectree}), 
for each cell $\D$, we add the total weight of the hyperplanes below $\D$ to the answer for each point in $\D$.
For each hyperplane $h$ below $\D$,
we also need to add the total weight of the points inside $\D$ to the answer for $h$.
Naively, this requires $O(nr^d)$ time, so (\ref{eqn:recur:dectree})
is weakened to
\[
 T'(m,n) \:=\: O(r^d)\, T'(m/r^d, n/r) + O(nr^d + m).
\]
Fortunately, this is still sufficient for the proofs of Lemmas \ref{lem:hopcroft:dectree}--\ref{lem:hopcroft:dectree2}, since the choices
of $r$ there are relatively small.
In the base case, we can use Lemma~\ref{lem:arrang}.

In the proof of Theorem~\ref{thm:hopcroft} and Corollary~\ref{cor:hopcroft}, we need to recover
(\ref{eqn:recur}) (so as to get (\ref{eqn:recur2}) and (\ref{eqn:logstar})).
To this end, we use the Hierarchical Cutting Lemma.
For each cell and each hyperplane $h$ that crosses the parent cell but is
completely below the cell, we can add the total weight of the points in the cell to the answer for $h$.  The cost is $O(\sum_{i=1}^k (\rho^i)^d\cdot n/\rho^{i-1})=O(nr^{d-1})$.
\end{myproof}

\Paragraph{Remark.}  The above result works in the semigroup setting, since we only need additions of weights.

\subsection{Shallow variant}

A similar approach can be applied to the problem of detecting
a point-above-hyperplane pair (a version of offline halfspace range
emptiness).  This time, we use a shallow version of the Cutting Lemma:

\begin{lemma}\label{lem:cut:shallow}
\emph{(Shallow Cutting Lemma)}
Given $n$ hyperplanes in a constant dimension~$d$ and a parameter $r\le n$,
there exists a cover of the region below the lower envelope by $O(r^{\down{d/2}})$ disjoint simplicial cells, such that each cell is
crossed by at most $n/r$ hyperplanes.  All cells are unbounded from below
(i.e., contain $(0,\ldots,0,-\infty)$).

The cells and the list of hyperplanes crossing each cell
can be constructed in $O(nr^{\down{d/2}-1})$ time.
Furthermore, given a set of $m$ points, we can find the cell containing each point (if it exists), in $O(m\log n)$ additional time.
\end{lemma}

A more general form of the Shallow Cutting
Lemma for the $(\le k)$-level, rather than just the lower envelope ($k=0$),
was originally stated by Matou\v sek~\cite{Matousek92CGTA}.
For the case $r\le n^{1-\eps}$, the above time bounds follow from
Matou\v sek's work~\cite{Matousek92CGTA} (which considered the case $r\le n^\alpha$ for
some small constant $\alpha$, but the extension to $r\le n^{1-\eps}$ 
can be handled by a constant number of rounds
of recursion).  
The stated time bounds hold for
all $r\le n$ according to Ramos's paper~\cite{Ramos99}, by using
techniques developed by
Br\"onnimann, Chazelle, and
Matou\v sek~\cite{BronnimannCM99}.%
\footnote{To find the cell containing each point, Ramos used ray shooting queries in convex polytopes, which require $O(r^{\down{d/2}}\log^{O(1)}r)$ preprocessing time (via a structure named ``$\mathcal{D}_1$'' in his paper).
This cost is absorbed by the $O(nr^{\down{d/2}-1})$ bound, provided that
$r\le n/
\log^{a'}n$ for some constant~$a'$ (which conveniently holds in our application).
Ramos later in his paper improved the polylogarithmic factors in 
the preprocessing time for ray shooting, so the result should hold
for all $r\le n$.
}
Again, simpler constructions are possible
if randomization is acceptable.
(A hierarchical version of the Shallow Cutting Lemma is known only for even dimensions, but fortunately is not needed in our algorithm.)

\begin{theorem}\label{thm:hopcroft:shallow}
Given $n$ points and $n$ hyperplanes in 
a constant dimension $d\ge 4$, we can decide the existence of a point-above-hyperplane
pair in $O(n^{2\down{d/2}/(\down{d/2}+1)})$ time.
\end{theorem}
\begin{myproof}
We follow exactly the same plan as in Section~\ref{sec:hopcroft:dectree}, but using the Shallow Cutting Lemma instead of the Cutting Lemma.  (If a point is not covered by the cells,
we know it is above some hyperplane and can terminate the algorithm.)
All occurrences of $d$ in the exponents are replaced by $\down{d/2}$.
\end{myproof}

\subsection{Online queries}

Hopcroft's problem is related to offline halfspace
range counting. In this section, we
show that our approach via 2D fractional cascading may
also be adapted to yield data structures for online
halfplane range counting queries in 2D.

\IGNORE{
\begin{theorem}
\emph{(Partition Theorem)}
Given a set $P$ of $n$ points in a constant dimension $d$
and a parameter $r\le n$, 
there exists a partition of $P$ into $r$ subsets
$P_1,\ldots,P_r$ each containing at most $n/r$ points,
and $r$ simplicial cells $\D_i\supset P_i$, such that
any hyperplane crosses at most
$O(r^{1-1/d})$ cells.  The partition and cells can
be constructed in $O(n\log n)$ time if $r\le n^{1-\eps}$
for some constant $\eps>0$.
\end{theorem}
}

We will need another standard tool for range searching: Matou\v sek's
Partition Theorem~\cite{Matousek92}.  The following strengthened, hierarchical version
was obtained by Chan~\cite{Chan12}:

\begin{theorem}
\emph{(Hierarchical Partition Theorem)}
Given $n$ points in a constant dimension $d$
and a parameter $s\le n$, for some constant $\rho>1$,
there exists a sequence $\Xi_1,\ldots,\Xi_k$ with $k={\up{\log_\rho s}}$,
where each $\Xi_i$ is a decomposition of $\R^d$ into $O(\rho^i)$ disjoint simplicial cells,  each cell contains
at most $n/\rho^i$ points, any hyperplane crosses at most
$O((\rho^i)^{1-1/d}+\log^{O(1)}n)$ cells of $\Xi_i$, and each cell in $\Xi_i$ is a union of $O(1)$
disjoint cells in $\Xi_{i+1}$.  All the cells can be constructed
in $O(n\log n)$ time w.h.p.\footnote{With high probability, i.e., $1-O(1/n^c)$ for an arbitrarily large constant $c$.}
\end{theorem}

We first review previous approaches to halfspace range counting.
One approach is to apply
the Hierarchical Partition Theorem, which in $O(n\log n)$ time w.h.p.\
generates a tree of cells with degree $O(1)$
and height $k=\up{\log_\rho r}$ (which
$\Xi_i$ corresponds to the cells at level $i$
of the tree).  For each leaf~$v$, let $P_v$ be the subset of all input points in $v$'s cell; the size of $P_v$ is at most $n/s$.
Suppose we have stored $P_v$
in a data structure with $P(n/s)$ preprocessing time and $Q(n/s)$ query time.
Given a query halfspace, 
we visit the cells in the tree crossed by its bounding hyperplane 
by proceeding top-down.
For each child cell of such cells that is completely inside the query halfspace, 
we add the number of points in the child cell to the counter.
The number of cells visited is $O(\sum_{i=1}^k (\rho^i)^{1-1/d} +\log^{O(1)}n)=O(s^{1-1/d}+\log^{O(1)}n)$.
Consequently, for any $s\in [\log^{\omega(1)}n, n]$, we obtain a new data structure with
(expected) preprocessing time and query time
\begin{eqnarray}
 P'(n)&=&O(s)P(n/s)+O(n\log n)\nonumber\\
 Q'(n)&=&O(s^{1-1/d})Q(n/s) +
O(s^{1-1/d}).\label{eqn:recur:online}
\end{eqnarray}

Another approach is to switch to the dual problem: counting the number
of hyperplanes below a query point for a given set of hyperplanes.
Apply the Hierarchical Cutting Lemma, which in $O(nr^{d-1})$ time
generates a tree of degree $O(1)$ and height $k=\up{\log_\rho r}$.
For each leaf~$v$, let $H_v$ be the subset of all
hyperplanes crossing $v$'s cell; the size of $H_v$ is at most $n/r$.
Suppose we have stored $H_v$ 
in a data structure with $P(n/r)$ preprocessing time and $Q(n/r)$ query time.
Given a query point, 
we visit the $O(\log r)$ cells in the tree containing the query point by proceeding top-down.
For each such cell,  we add the number of hyperplanes that cross its parent cell but are also completely below the cell, to the counter.
Consequently, for any $r\le n$, we obtain a new data structure with
preprocessing and query time
\begin{eqnarray}
 P''(n)&=& O(r^d)P(n/r)+O(nr^{d-1})\nonumber\\
 Q''(n) &=& Q(n/r) +
O(\log r). \label{eqn:recur:online2}
\end{eqnarray}

By applying (\ref{eqn:recur:online}) and (\ref{eqn:recur:online2}) in succession,
and setting $s=r^d$, we
get another data structure with preprocessing and query time
\begin{eqnarray*} 
 P'''(n)&=& O(r^{2d})P(n/r^{d+1})+O(n\log n + nr^{d-1})\\
 Q'''(n)&=& O(r^{d-1})Q(n/r^{d+1}) +
O(r^{d-1}\log r),
\end{eqnarray*}
assuming that $r\le n/r^d$ and $r^d\ge \log^{\omega(1)}n$.
By choosing $r=(n/\log^C n)^{1/(d+1)}$ for a sufficiently large constant~$C$,
\begin{eqnarray}
 P'''(n) &=& O(n/\log^C n)^{2d/(d+1)} P(\log^C n) + o(n^{2d/(d+1)})\nonumber\\
 Q'''(n) &=& O(n/\log^C n)^{(d-1)/(d+1)} Q(\log^C n) + 
o(n^{(d-1)/(d+1)}).\label{eqn:logstar:online}
\end{eqnarray}
By recursion, this gives a data structure with preprocessing time $\;$ $n^{2d/(d+1)}2^{O(\log^*n)}$
and query time $n^{(d-1)/(d+1)}2^{O(\log^*n)}$, a result obtained
by Chan~\cite{Chan12}.

We show how to eliminate the $2^{O(\log^*n)}$ factor in the
2D case, by adapting our approach for fractional cascading in
arrangements of lines.

First, notice that Lemma~\ref{lem:cascade} can already
handle queries online.  The key issue, however, is the assumption of $x$-monotone queries.  We show
how to remove this assumption (and also avoid amortization)
by switching to the decision tree setting, counting only the cost of comparisons.

\begin{lemma} \label{lem:cascade:online}
There is a randomized data structure for Problem~\ref{prob:cascade} which can be preprocessed using 
$O(|T|z^2+n\log n)$ expected number of comparisons, such that a query
for a fixed point $q$ and subtree $T_q$ can be answered 
using $O(\log(|T|z) + |T_q|)$ expected number of comparisons,
assuming that all sets $L_v$ are subsets of a common set $L$
of $n$ lines.
\end{lemma}
\begin{myproof}
We modify the proof of Lemma~\ref{lem:cascade}.
In the preprocessing, let $X$ be the set of $x$-coordinates of 
the $O(|T|z^2)$ vertices from all the arrangements in step~1. 
We first sort $X$.  By Fredman's sorting technique~\cite{Fredman76}, this can be
done using $O(|T|z^2 + n\log n)$ comparisons
(e.g., see Appendix~\ref{app:dectree}, or if a weaker 
$O(|T|z^2 + n^{1+\eps})$ bound suffices, see Section~\ref{sec:sort}).

In a query for the dual point $q$, we first perform a predecessor
search for its $x$-coordinate among the values in $X$, in $O(\log(|T|z))$
comparisons.  The part of the query algorithm that requires
$x$-monotonicity (and amortization) is step~(ii), but step~(ii)
reduces to predecessor search of the $x$-coordinate of $q$
among the $x$-coordinates of the
vertices of $f'$, and does not require any new comparisons,
since we already know the rank of $q$ in $X$.
\end{myproof}

\begin{lemma}\label{lem:dectree:online}
There is a randomized data structure for $n$ points
in $\R^2$ which can be preprocessed using
$O(n^{4/3})$ expected number of comparisons, such that
a halfspace range counting query
can be answered using $O(n^{1/3})$ expected number of comparisons.
\end{lemma}
\begin{myproof}
We apply the Hierarchical Partition Theorem in the same way that
we derived (\ref{eqn:recur:online}), but we will use Lemma~\ref{lem:cascade}
to handle the subproblems at the leaves.
More precisely, we create an instance of Problem~\ref{prob:cascade}
for a tree $T$ with $|T|=O(s)$ nodes,
where the set of lines at each leaf node is the dual set $P_v^*$, which has size at most
$z := n/s$,
and the set of lines at each internal node is the empty set.

Suppose we are given a query upper halfplane bounded by a line $\ell$.
For each leaf cell~$v$ crossed by $\ell$, the subproblem of counting the number of points in $P_v$ above $\ell$ can be solved by finding the face of $\A(P_v^*)$ containing the
dual point $\ell^*$.  This corresponds to a query for Problem~\ref{prob:cascade} for the subtree $T_{\ell^*}$ of all nodes 
whose cells are crossed by~$\ell$---these nodes indeed form
a subtree containing the root, since if a cell is crossed by $\ell$,
all ancestor cells are crossed by $\ell$.  
The size of $T_{\ell^*}$ is bounded by
$O(\sum_{i=1}^k \sqrt{\rho^i} + \log^{O(1)}n) = O(\sqrt{s}+\log^{O(1)}n)$.

By Lemma~\ref{lem:cascade:online},
preprocessing requires $O(|T|z^2+n\log n)=O(s(n/s)^2+n\log n)$ expected number of
comparisons, and a query requires $O(\log(|T|z) + |T_{\ell^*}|)=
O( \sqrt{s} + \log^{O(1)} n)$ expected number
of comparisons.  Choose $s=n^{2/3}$.
\end{myproof}

As before, we can convert decision tree complexity to time complexity:

\begin{theorem}\label{thm:online}
There is a randomized data structure for $n$ points
in $\R^2$ with
$O(n^{4/3})$ expected preprocessing time and space, such that
a halfspace range counting query 
can be answered in $O(n^{1/3})$ expected time.
\end{theorem}
\begin{myproof}
Applying (\ref{eqn:logstar:online}) three times for $d=2$ gives
\begin{eqnarray}
 P'''(n) &=& O(n/b)^{4/3} P(b) + o(n^{4/3})\nonumber\\
 Q'''(n) &=& O(n/b)^{1/3} Q(b) + o(n^{1/3})\label{eqn:final:online}
\end{eqnarray}
with $b=O((\log\log\log n)^C)$.

Since the new input size $b$ is tiny, we can afford to build
the decision trees from Lemma~\ref{lem:dectree:online} in advance, and
get $P(b)=O(b^{4/3})$ and $Q(b)=O(b^{1/3})$. 

(Viewed another way: we can directly simulate the algorithm from Lemma~\ref{lem:dectree:online} when the input size $b$ is tiny,
by using bit packing and table lookup.)

As a result, (\ref{eqn:final:online}) implies $P'''(n)=O(n^{4/3})$
and $Q'''(n)=O(n^{1/3})$.
\end{myproof}

\Paragraph{Remarks.}
The time bounds above actually hold with high probability ($1-2^{-n^{\Omega(1)}}$) by applying the Chernoff bound,
since $P'''(n)$ and $Q'''(n)$ above are both
sums of many independent random variables
(if we use a fresh set of random choices for each subproblem of size $b$).
Since there are only polynomially many ``different'' query halfspaces
(with respect to the predicates used by the algorithm), we also
obtain a high-probability bound for the \emph{maximum} query time,
by a union bound.  In particular, we no longer need to assume that the queries
are oblivious to the random choices made by the data structure.

A complete range of trade-offs between preprocessing and query time follows:

\begin{corollary}\label{cor:tradeoff}
For any $M\in [n\log n,\, n^2/\log^2n]$,
there is a randomized data structure for $n$ points
in $\R^2$ with $O(M)$ expected preprocessing time and space, such that
a halfspace range counting query 
can be answered in $O(n/\sqrt{M})$ expected time.
\end{corollary}
\begin{myproof}
If $M\le n^{4/3-\eps}$,
put $P(n/s)=O((n/s)^{4/3})$
and $Q(n/s)=O((n/s)^{1/3})$ into (\ref{eqn:recur:online}).
Then $P'(n)=O(s(n/s)^{4/3}+n\log n)$ and $Q'(n)=O(\sqrt{s}(n/s)^{1/3})$.
Set $s$ so that $s(n/s)^{4/3}=M$.

If $M > n^{4/3-\eps}$,
put $P(n/r)=O((n/r)^{4/3-\eps})$ and $Q(n/r)=O((n/r)^{1/3+\eps/2})$ into (\ref{eqn:recur:online2}).
Then $P''(n)=O(r^2(n/r)^{4/3-\eps})$ and $Q''(n)=O((n/r)^{1/3+\eps/2}+\log r)$.
Set $r$ so that $r^2(n/r)^{4/3-\eps}=M$.
\end{myproof}

The above result generalizes to summing the weights of the points inside a query halfplane, in the semigroup setting.  The trade-offs match precisely Chazelle's lower bound in the semigroup model~\cite{Chazelle89}.

\section{Applications}\label{sec:app}

In this section, we describe just some of the numerous
consequences of our new results. 

\subsection{Triangle range counting, line segment intersection counting,
and ray shooting}

\begin{corollary}\label{cor:segint}\
\begin{enumerate}
\item[(a)] 
There is a randomized data structure for $n$ points $p_1,\ldots,p_n$
in $\R^2$ with
$O(n^{4/3})$ expected preprocessing time and space, such that
the number of points in any prefix $p_1,\ldots,p_i$ inside a query halfplane
can be counted in $O(n^{1/3})$ expected time.
\item[(b)] 
There is a randomized data structure for $n$ points
in $\R^2$ with
$O(n^{4/3})$ expected preprocessing time and space, such that
the number of points inside a query triangle 
can be counted in $O(n^{1/3})$ expected time.
\item[(c)] There is a randomized data structure for $n$ triangles
in $\R^2$ with
$O(n^{4/3})$ expected preprocessing time and space, such that
the number of triangles containing a query point
can be counted in $O(n^{1/3})$ expected time.
\item[(d)] There is a randomized data structure for $n$ line segments
in $\R^2$ with
$O(n^{4/3})$ expected preprocessing time and space, such that
the number of line segments intersecting a query line segment
can be counted in $O(n^{1/3})$ expected time.
\item[(e)] There is a randomized data structure for $n$ line segments
in $\R^2$ with
$O(n^{4/3})$ expected preprocessing time and space, such that
one line segment intersecting a query line segment (if it exists)
can be reported in $O(n^{1/3})$ expected time.
\item[(f)] There is a randomized data structure for $n$ line segments
in $\R^2$ with
$O(n^{4/3})$ expected preprocessing time and space, such that
a ray shooting query can be answered in $O(n^{1/3})$ expected time.
\end{enumerate}
\end{corollary}
\begin{myproof}\
\begin{enumerate}
\item[(a)] This follows by
straightforward divide-and-conquer: build a halfplane range
counting structure for $p_1,\ldots,p_n$, and recursively
build the data structures for $p_1,\ldots,p_{n/2}$ and for $p_{n/2},\ldots,p_n$.
Using the halfplane range counting structure
from Theorem~\ref{thm:online},
the new data structure has expected preprocessing
time $P(n)=2P(n/2)+O(n^{4/3})$ and query time $Q(n)=Q(n/2)+O(n^{1/3})$,
implying $P(n)=O(n^{4/3})$ and $Q(n)=O(n^{1/3})$.

\item[(b)]  By additions and subtractions,
the problem reduces to counting the number of points
above a query line segment, which in turn reduces to counting
the number of points above a query leftward ray.
This is an instance of (a), if we order the points by $x$-coordinates.  (The idea of using subtractions to reduce the triangle case to line segments was also used, for example, in Agarwal's previous algorithm~\cite{Agarwal90ii}.)

\item[(c)]   By additions and subtractions,
the problem reduces to counting the number of line segments
below a query point, which in turn reduces to counting
the number of rightward rays below a query point.
If we order the rays by $x$-coordinates
of their endpoints, the problem reduces to
counting the number of lines below a query point,
among a prefix of a given sequence of lines;
in the dual, this is an instance of (a).

\item[(d)] 
First consider the case when the input line segments are lines.  By duality, counting the number of lines intersecting
a query line segment reduces to counting the number of points
inside a query double-wedge, which is an instance of (a).

Next consider the ``opposite'' case when the query line segment is a line.
By duality, counting the number of line segments intersecting
a query line reduces to counting the number of double-wedges
containing a query point, which is an instance of (b).

Finally consider the general case. We use divide-and-conquer in the
style of a segment tree~\cite{ChazelleEGS94,BergCKO08}:
Suppose all line segments are in
a vertical slab $\sigma$.  Call a segment \emph{long}
if both endpoints lie on the boundary of $\sigma$, and \emph{short} otherwise.
Build a data structure from (a) to handle the case of long input
segments, and a data structure from (b) to handle the case of
long query segments.
For the remaining short input
segments, divide $\sigma$ into two subslabs by the median $x$-coordinate
of the endpoints in the interior of $\sigma$, and recursively build the 
data structures for the two subslabs.
The preprocessing time satisfies $P'(n)=2P'(n/2)+O(n^{4/3})$, implying 
$P'(n)=O(n^{4/3})$.

To answer a query for a ray, we first query the structure from (a)
to handle all the long input segments.  For the remaining
short input segments,
we recursively query in the subslab containing the ray's endpoint, and
if the ray intersects the other subslab, we query the structure from (b)
(as the ray would become a long segment in the other subslab).
The query time satisfies $Q'(n)=Q'(n/2)+O(n^{4/3})$, implying
$Q'(n)=O(n^{1/3})$.  

By subtraction, a query for a line segment reduces to two queries for rays.

\item[(e)] Build the data structure from (d), arbitrarily
divide the input set into two halves, and recursively build the
data structures for two halves.
The preprocessing time satisfies $P''(n)=2P''(n/2)+O(n^{4/3})$, implying 
$P''(n)=O(n^{4/3})$.
To answer a query, we query the structure from (d) on the first half,
and if the count is nonzero, we recursively query the first half, else
we recursively query the second half.
The query time satisfies $Q''(n)=Q''(n/2)+O(n^{1/3})$, implying
$Q''(n)=O(n^{1/3})$.

\item[(f)] The problem reduces to the decision problem (deciding
whether a query line segment intersects any input line segment)
by a known randomized optimization technique~\cite{Chan99}, and
the decision problem can be solved by the data structure from (d).  The
reduction increases the preprocessing and query time only by constant factors if
the preprocessing time exceeds $n^{1+\eps}$ and the query time
exceeds $n^\eps$.
\end{enumerate}

\vspace{-\bigskipamount}
\end{myproof}

Trade-offs between preprocessing and query time for the above
data structures are also possible, similar to Corollary~\ref{cor:tradeoff}, at least for
$M\in [n^{1+\eps},n^{2-\eps}]$.
For Corollary~\ref{cor:segint}(a)--(e), we can also generalize counting to summing weights in the group (but not the semigroup) model.

Corollary~\ref{cor:segint}(f) in particular implies the following.
(Since offline queries are sufficient here, Theorem~\ref{thm:online} and the use of the Partition Theorem
are not necessary, and the resulting algorithm
is deterministic by Theorem~\ref{thm:hopcroft}.)

\begin{corollary} \label{cor:offline:segint}
Given $n$ line segments in $\R^2$,
we can count the number of intersections in
$O(n^{4/3})$ time.
\end{corollary}

\subsection{Connected components of line segments}

We describe an application to another problem about line segments: 

\begin{corollary} \label{cor:ConnectedComponents}
  Given $n$ line segments, we can report the connected components in $O(n^{4/3})$ time.
\end{corollary}
\begin{myproof}
We will generate a sparse graph $G$, with the input segments as vertices, and with $O(n\log n)$ edges, so that two segments are connected iff they are in the same connected component in $G$.
Afterwards, we can easily run a graph traversal algorithm (e.g., depth-first search) on $G$ to report the connected components.

To this end, we use divide-and-conquer in the style of a segment tree, like we did earlier for segment intersection queries in the proof of Corollary~\ref{cor:segint}(d).

Consider a vertical slab~$\sigma$. 
We first compute the connected components of the 
long segments
in $O(n\log n)$ time, as follows
(a similar subroutine was also used in the previous algorithm by Lopez and Thurimella~\cite{LopezT95}).
We go through the 
segments in decreasing $y$-coordinates of 
left endpoints, and maintain a stack $S$ of 
components 
sorted 
by the lowest right endpoint of the component.
Each segment we consider either forms 
a new component and is added on top of the stack $S$, or is
merged with one or more components on top of the stack.
If the long segment goes through multiple components, 
we merge those into one component. 

Having computed the components of the long segments within $\sigma$, we add 
an arbitrary spanning tree of each component to $G$; the number of edges added is $O(n)$.
We also compute the lower and upper envelopes of each connected component in $O(n\log n)$ total time.
These envelopes are disjoint, and for each short segment $s$, we can do point location among these envelopes for the 
endpoints of $s$ in $O(\log n)$ time~\cite{Kirkpatrick83}.
We consider two cases:
\begin{enumerate}
\item[(a)] If the endpoints of $s$ lie within different envelopes,
we know exactly which components the 
short segment $s$ intersects (see Figure~\ref{fig:my_label}(a)).  Among these components, we take each pair of
consecutive components and add an edge between two representative long segments of
the two components to $G$, if the edge was not added before.
We also add an edge from $s$ to a representative long segment in one such component to $G$.
\item[(b)]
Otherwise, $s$ lies between an upper and lower envelope of a single connected component (see Figure~\ref{fig:my_label}(b)), so we can 
use Corollary~\ref{cor:segint}(e) to decide 
whether $s$ intersects 
that component, and if so, report one long segment intersected by $s$,
for all short segments $s$, in $O(n^{4/3})$ total time.
(Actually we only need Corollary~\ref{cor:segint}(e)
for the case when the input segments are lines.)
As offline queries are sufficient,
this can be made deterministic by
Theorem~\ref{thm:hopcroft}.
We add an edge from $s$ to the reported long segment intersected by $s$ (if it exists) to $G$.  
\end{enumerate}

The number of edges added to $G$ is $O(n)$.  We now remove the long segments.
To handle connectivity among the remaining short 
segments, divide $\sigma$ into two subslabs by the median $x$-coordinate
of the endpoints in the interior of $\sigma$, and make recursive calls for the two subslabs.  It is straightforward to verify by induction that the resulting graph $G$ preserves the connectivity of the input segments.
The running time satisfies the recurrence
$T(n)=2T(n/2)+O(n^{4/3})$ (where $n$ here denotes the number of endpoints
in the interior of $\sigma$), and the number of edges added to $G$ satisfies
the recurrence $E(n)=2E(n/2)+O(n)$.
These recurrences yield $T(n)=O(n^{4/3})$ and $E(n)=O(n\log n)$.
\begin{figure}
    \centering
    \begin{subfigure}{.45\textwidth}
    \centering
    \includegraphics[page=1, scale=0.4]{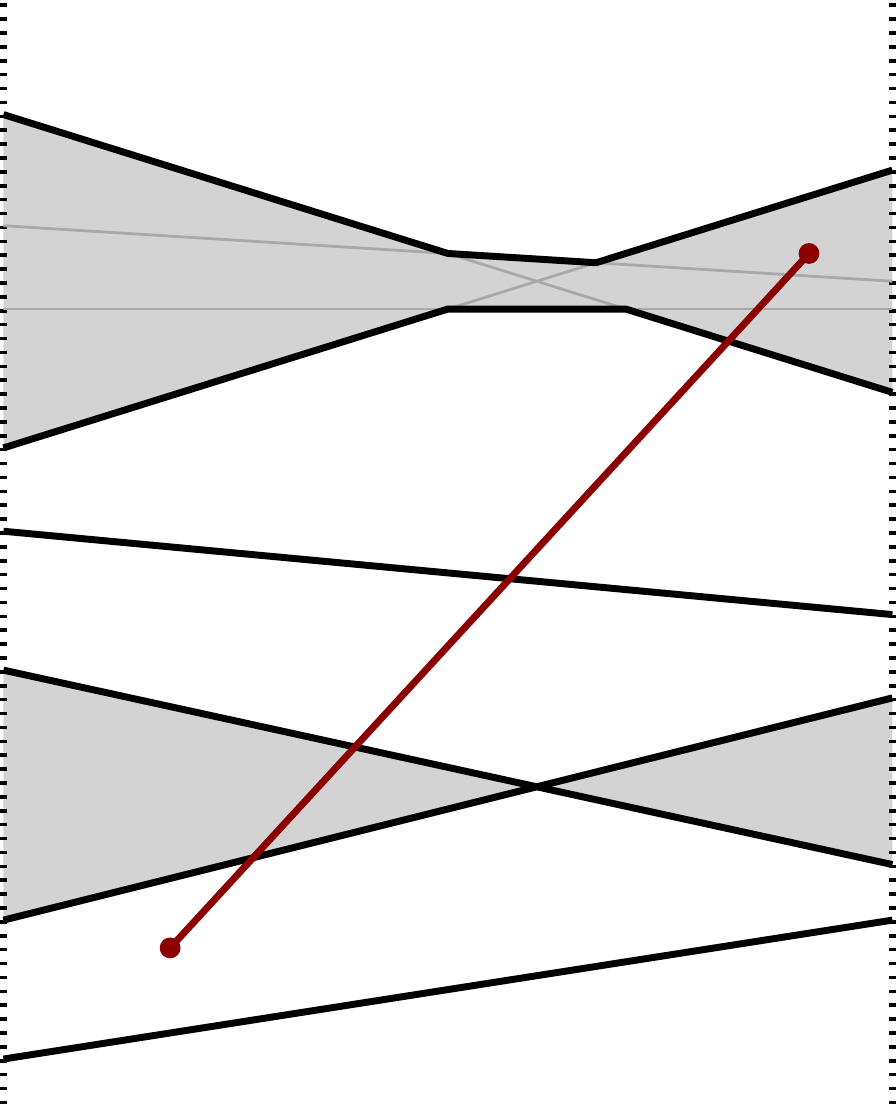}
    \caption{}
    \end{subfigure}
    \begin{subfigure}{.45\textwidth}
    \centering
    \includegraphics[page=2, scale=0.4]{slab_connected_comp2.pdf}
    \caption{}
    \end{subfigure}
    \caption{
    (a) The endpoint of the short segment 
    lies between different upper and lower envelopes of connected components.
    (b) The endpoints lie within 
    the upper and lower envelope of one component,
    so we need to check if the short segment 
    intersects any of the long segments of that component.}
    \label{fig:my_label}
\end{figure}
\end{myproof}

\subsection{Bichromatic closest pair, Euclidean minimum spanning tree, and more}

\begin{corollary}\
\begin{enumerate}
\item[(a)]
Given $n$ red points and $n$ blue points in 
a constant dimension $d\ge 3$, we can decide the
existence of a bichromatic pair of distance at most $1$
in $O(n^{2\up{d/2}/(\up{d/2}+1)})$ time.
\item[(b)]
Given $n$ red points and $n$ blue points in 
a constant dimension $d\ge 3$, we can find the bichromatic closest pair
in $O(n^{2\up{d/2}/(\up{d/2}+1)})$ expected time.
\item[(c)]
Given $n$ points in 
a constant dimension $d\ge 3$, we can find the Euclidean minimum spanning tree 
in $O(n^{2\up{d/2}/(\up{d/2}+1)})$ expected time.
\end{enumerate}
\end{corollary}
\begin{myproof}
By a standard lifting transformation, (a)
reduces to deciding the existence of a point-above-hyperplane pair for
$n$ points and $n$ hyperplanes in $\R^{d+1}$, and Theorem~\ref{thm:hopcroft:shallow} yields an
$O(n^{2\up{d/2}/(\up{d/2}+1)})$-time algorithm.

(b) reduces to (a) by Chan's randomized
optimization technique~\cite{Chan99}. The reduction increases
the time bound only  
by a constant factor.  

(c) reduces to (b)
by a method of Agarwal et al.~\cite{AgarwalES91}.
The reduction increases the time bound only by a constant factor if it exceeds $n^{1+\eps}$.
\hfill
\end{myproof}

\begin{corollary}\
\begin{enumerate}
\item[(a)]
Given $n$ red lines and $n$ blue lines in $\R^3$, we can solve the \emph{line towering problem}, i.e.,
decide whether some red line is above some blue line,
in $O(n^{4/3})$ time.

\item[(b)]
Given $n$ red lines and $n$ blue lines in $\R^3$ where all the red lines are below all the blue lines, we can find the bichromatic pair of lines with the smallest vertical distance in $O(n^{4/3})$ expected time.
\item[(c)]
Given two nonintersecting polyhedral terrains of size $n$ in 
$\R^3$, we can find the minimum vertical distance in $O(n^{4/3})$ expected time.
\end{enumerate}
\end{corollary}
\begin{myproof}
(a) reduces to deciding the existence of a point-above-hyperplane pair for
$n$ points and $n$ hyperplanes in $\R^5$, as shown by Chazelle et al.~\cite{ChazelleEGSS96}, using
Pl\"{u}cker coordinates.
The reduction increases the time bound only by a constant factor if it exceeds $n^{1+\eps}$.
Theorem~\ref{thm:hopcroft:shallow} thus implies an $O(n^{4/3})$-time algorithm.

(b) reduces to (a) by Chan's randomized
optimization technique~\cite{Chan99}. 

(c) reduces to (b) by a method of Chazelle et al.~\cite{ChazelleEGS94}, which
uses a two-level hereditary segment tree.  With some minor changes
to their algorithm and analysis,
the reduction increases the time bound only by a constant factor
when it exceeds $n^{1+\eps}$.  
(Vaguely: the time bound for the first-level
subproblems 
satisfies the recurrence $T_1(n)=2T_1(n/2)+O(T_0(n))$, where
$T_0(n)$ is the complexity of the problem in (b), and the
time bound for the second-level subproblems satisfies the recurrence
$T_2(n)=2T_2(n/2)+O(T_1(n))$.  For $T_0(n)=O(n^{4/3})$,
these recurrences solve to $T_1(n)=T_2(n)=O(n^{4/3})$.)  
\hfill
\end{myproof}

\subsection{Distance ranking and selection}

\begin{corollary}
Given $n$ red points and $n$ blue points in $\R^2$, 
\begin{enumerate}
    \item[(a)] we can count the number of bichromatic pairs
that have distance at most~$1$, and more generally count the number of blue points with distance at most~1 from each red point, in
$O(n^{4/3})$ time;
\item[(b)] we can select the $k$-th smallest distance among all bichromatic pairs of points for a given $k$ in $O(n^{4/3})$  time w.h.p.
\end{enumerate}
\end{corollary}
\begin{myproof}\ 

\begin{enumerate}
    \item[(a)]
The problem is equivalent to: given $n$ red points and $n$ blue unit circles in
$\R^2$, count the number of red points inside each blue circle and the number of blue circles containing each red point.  
The cutting lemma still holds with lines replaced by  unit circles in the plane,
and it is straightforward to adapt the algorithm 
for 2D Hopcroft's problem in Section~\ref{sec:dectree} to solve this problem. (More details can also be found in
a recent paper by Wang~\cite{Wang22}.)  Note that like Hopcroft's, the problem here is its own dual (by switching red and blue points).

\item[(b)]
We adapt a standard selection algorithm by Floyd and Rivest via random sampling~\cite{FloydRivest}.  To select the $k$-th smallest element in a set $X$ of size $m$, one version of the algorithm proceeds as follows, for given parameters $s$ and $g$:
\begin{enumerate}
    \item[1.] Let $S$ be a (multi)set of $s$ randomly chosen elements from $X$. 
    \item[2.] Find the $\lceil ks/m - g\rceil$-th smallest element $a$ and the $\lceil ks/m + g\rceil$-th smallest element of $S$.
    \item[3.] Compute the rank $k_a$ of $a$ in $X$ and the rank $k_b$ of $b$ in $X$.
    \item[4.] If $k_a< k\le k_b$, then  return the $(k-k_a)$-th smallest element of $X\cap (a,b]$, else declare failure.
\end{enumerate}
(If $ks/m-g\le 0$, redefine $a=-\infty$ and $k_a=0$.
If $ks/m+g > s$, redefine $b=\infty$ and $k_b=m+1$.)
By known analysis~\cite{Kiw} (see also~\cite{MatIPL91}),  step~4 succeeds and $|X\cap (a,b]| = O(gm/s)$ with probability $1-e^{-\Omega(g^2/s)}$.

In our application, $X$ is the set of distances of all bichromatic pairs lying in a given interval $(a_0,b_0]$ (initially, all of $\R$); here, $m=O(n^2)$.
We choose $s=n^{0.3}$ and $g=n^{0.2}$.
However, we cannot afford to explicitly generate $X$.  

To implement step~1 efficiently, we first compute, for each red point $p$, the number of blue points of distance at most $a_0$ from $p$ and the number of blue points of distance at most $b_0$ from $p$, by two calls to the algorithm in part~(a) (with appropriate rescaling) in $O(n^{4/3})$ time. By taking the difference, we also obtain the number $c_p$ of blue points of distance in $(a_0,b_0]$ from $p$.  Note that $\sum_p c_p = |X|$.  To generate one random element of $X$, we randomly choose a red point $p$ according to the distribution in which $p$ is chosen with probability $c_p/|X|$; we then naively enumerate all distances of the blue points from $p$ in $O(n)$ time and choose a random distance that lies in $(a_0,b_0]$.  Thus,
step~1 (generating $s$ random elements of $X$) can be implemented
in $O(n^{4/3}+sn)\le O(n^{4/3})$ time.

Step~2 takes $O(s)$ time.  Step~3 can be done by the algorithm in part~(a) in $O(n^{4/3})$ time.  Step~4 can be done recursively, with $(a_0,b_0]$ replaced by $(a_0,b_0]\cap (a,b]$.
Since the number of elements $m$ is reduced to $O(gm/s)\le O(m/n^{0.1})$ w.h.p., the number of rounds of recursion is $O(1)$ and
the total running time is $O(1)$ w.h.p.  (We can switch to a slow algorithm if failure is encountered.)
\end{enumerate}

\vspace{-1.5\bigskipamount}
\end{myproof}

Note that the more standard  ``monochromatic'' distance selection problem easily reduces to the bichromatic version above.

\IGNORE{ 

\TIMOTHY{other potential problems, which might require nontrivial extra work:

biclique cover?? maybe not...

distance ranking

distance selection
(I think it's doable: sampling distances
in an interval reduces to counting per point?)

k-sensitive distance ranking/selection...

offline halfspace range emptiness where we want individual answers for each halfspace (this might need shallow partition trees, so only even $d$?)

offline halfspace range reporting? other reporting problems?
(e.g., point-line pairs of vertical distance at most 1?)

extreme points?

multiple faces in arrangements of lines? (Agarwal et al.)

circle or arc intersection counting in $n^{3/2}$? (Agarwal and Sharir)

offline simplex range counting in higher dimensions: seems doable?

offline triangle range sum for semigroup?

verifying depth order?
}

\DAVID{
verifying depth order involves 
ray shooting among curtains 
(downward half-plane beneath a line in 3d)

this can reduce to general ray shooting among half-planes 
which \cite{AgarwalM93} solve with a 2 level data structure:

- the first level is a partition tree

- the second level is the line towering problem Corollary 6.5(a)

}

}

\section{More Observations on Decision Trees}\label{app:dectree}

Returning to the decision tree approach from Section~\ref{sec:dectree},
we give an alternative to
the Basic Search Lemma (Lemma~\ref{lem:choice}), which can be viewed
as a generalization of Fredman's original technique~\cite{Fredman76}:

\begin{lemma}\label{lem:PL}
\emph{(Search Lemma---Second Version)}
Consider the $\Gamma$-algorithm framework.
Let $G$ be a dag with maximum out-degree $b$,
where each node $v$ is associated with a predicate $\gamma_v\in\Gamma$. Suppose
we are promised that for each non-sink node $v$ and for every input $\xx$ in the active cells,
$\gamma_v(\xx)$ implies $\gamma_{v'}(\xx)$ for some out-neighbor $v'$ of~$v$. 
Suppose we are also promised that  for some source node $s$, for every input $\xx$ in the active cells, $\gamma_s(\xx)$ is true.  Then
we can search for a sink $t$ such that $\gamma_t(\xx)$ is true by making
$O(1 - b\Delta\Phi)$ $\Gamma$-comparisons.
\end{lemma}
\begin{myproof} 

Define the \emph{weight} of a node $v$ to be 
the number of active cells satisfying $\gamma_v$.
Assume that the nodes are ordered so that the edges point downward.
Pick a lowest node $v$ in $G$ with weight at least $\frac1{b+1}
|\Act|$ (this is analogous to a weighted tree centroid, if the dag is a tree).  Each out-neighbor $v'$ of $v$
has weight at most $\frac 1{b+1}|\Act|$.
Make the comparison defined by $\gamma_v$.

\begin{itemize}
\item {\sc Case 0:} $\gamma_v(\xx)$ is true and $v$ is a sink.  We are done.
\item {\sc Case 1:} $\gamma_v(\xx)$ is true and $v$ is not a sink.
Active cells now satisfy $\gamma_v$, and thus satisfy $\gamma_{v'}$ for at least one of the $b$ out-neighbors $v'$ of $v$.  So, the number of active cells after the comparison is 
at most $\frac b{b+1}|\Act|$.
\item {\sc Case 2:} $\gamma_v(\xx)$ is false.
Active cells now do not satisfy $\gamma_v$.
So, the number of active cells after the comparison is  
at most $\frac b{b+1}|\Act|$.
\end{itemize}

In Cases 1 and 2, the potential $\Phi$ thus decreases by at least $\log \frac {b+1}b = \Omega(\frac1b)$.  Now repeat (with a new assignment of weights).  
The number of iterations is bounded by $O(-b\Delta\Phi)$.  
\end{myproof}

For example, we can find the predecessor of a value among a list of $M$
sorted values using $O(1-\Delta\Phi)$ $\Gamma$-comparisons, by applying the
above lemma to a binary tree with $M$ leaves and $b=2$.  Consequently, 
we can improve Theorem~\ref{thm:sort} to sort
$M$ values using $O(M-\Delta\Phi)\le O(M+N\log N)$ comparisons---this is
Fredman's sorting result.

Note that the algorithm is just a form of 
weighted binary search
(for example, commonly seen in distribution-sensitive data structures~\cite{BoseHM13}, whose analyses also often involve telescoping sums of logarithms).  However, what is daring about Fredman's technique is
that the weighted search is done in a very large space (cells in an $N$-dimensional arrangement).

As an original application, our generalized technique implies the following result about point location in multiple planar
subdivisions (which normally needs $O(L+M\log L)$ time):

\begin{theorem}
Given $t$ planar triangulated subdivisions $S_1,\ldots,S_t$ of total size $L$, and $t$ (not necessarily disjoint) planar point sets $Q_1,\ldots,Q_t$ of total size $M$, 
we can find the cell of $S_i$ that contains each query point $q\in Q_i$, for each $i\in [t]$,
using a total of $O(L+M-\Delta\Phi)\le O(L+M+N\log N)$ $\Gamma$-comparisons,
assuming that certain primitive operations on the vertices/edges of the $S_i$'s and points of the $Q_i$'s
can be expressed as $\Gamma$-comparisons on the original input $\xx\in\R^N$, and assuming that $\Gamma$ is reasonable.
\end{theorem}
\begin{myproof}
Store each subdivision 
in Kirkpatrick's point location structure~\cite{Kirkpatrick83}, which can be built in 
linear ($O(L)$) time.
Since Kirkpatrick's structure is a dag of out-degree $b=O(1)$, where
each node corresponds to a triangular cell and the sinks correspond to the cells of the subdivision,
we can find the cell containing a query point using just $O(1-\Delta\Phi)$ $\Gamma$-comparisons by Lemma~\ref{lem:PL}.  The total cost of the $M$ point location queries
is $O(M-\Delta\Phi)$.
\end{myproof}

It is amusing to compare the above result with the 
lower bounds of Chazelle and Liu~\cite{ChazelleL04}, which supposedly rule out
such improvements in the pointer
machine model for simultaneous planar point location (even in the simplest case when each subdivision is formed by just a set of parallel lines\footnote{
In the dual, this case is equivalent to the following: given $t$ vertical lines in the plane, where each line is decomposed into vertical line segments,
find the $t$ line segments intersecting a query line.
}).  The key differences are that (i)~we are considering
decision tree instead of time complexity, and (ii)~we
are working with offline instead of online queries (or more crucially,
we don't care about the worst-case cost of every query but may amortize cost).

There are a number of problems in computational geometry that
can be solved by point location in multiple planar subdivisions that
originate from a common input set.  For example, suppose we are given $n$ input points in $\R^2$, each with a time value.  We want to answer the following type of queries: find the nearest neighbor to a query point among those input 
points with time values in a query interval.  A standard solution is to
build a tree of Voronoi diagrams of various subsets, of total size $O(n\log n)$, which requires  $O(n\log n)$ preprocessing
time (using a known linear-time merging procedure for Voronoi diagrams~\cite{Kirkpatrick79}).  A query can then be answered by performing point location on $O(\log n)$ nodes
of the tree, yielding $O(\log^2n)$ time.  The lack of a 2-dimensional
analog of fractional cascading makes it difficult to improve the query time.
The best time bound known for answering $n$ such queries is $O(n\log^2n/\log\log n)$, obtained by slightly increasing the fan-out of the tree to $\log^\eps n$.
The above theorem (with $L,M=O(n\log n)$ and $N=O(n)$) immediately implies that a batch of $n$ queries can be answered using $O(n\log n)$ comparisons!
Unfortunately, for this particular problem, improved decision trees do not translate
to improved time complexity.  Nevertheless, the result rules out $\omega(n\log n)$
lower bounds in the algebraic decision tree model (which is arguably the most
popular model for proving lower bounds in computational geometry).

There appear to be many more examples of problems where we could
similarly shave log factors when working with decision trees.
For example, Seidel's $O(n^2+f\log n)$-time convex hull algorithm~\cite{Seidel86}
(where $f$ denotes the output size) can be improved to give
a decision tree of $O(n^2+f)$ depth.  Again, this does not appear to lead
improved running times.


\section{Final Remarks}

A few open questions remain.  For example, could
our data structure for online halfplane range counting queries in 2D be derandomized?
Could similar results be obtained for
online halfspace range counting queries in 3D and higher?

The most intriguing open question is whether the algebraic decision tree complexity of Hopcroft's problem could actually be near linear.  If so, one could then get
algorithms with polylogarithmic speedup over $n^{4/3}$.  In view of 
the known $\Omega(n^{4/3})$ lower bounds in restricted settings~\cite{Erickson96,Chazelle97}, one might think that the answer is no.  However, in 
a surprising breakthrough,
Kane, Lovett, and Moran~\cite{KaneLM19} recently obtained near-linear upper bounds
on the decision tree complexity for a class of problems including 3SUM and
$X+Y$ sorting.  So a yes answer is not completely out of the question.  (On the other hand,
Kane et al.'s result is about
point location in $O(n)$-dimensional arrangements of $n^{O(1)}$ hyperplanes with bounded
integer coefficients, while
Hopcroft's problem reduces to point location
in a certain $O(n)$-dimensional arrangement of $O(n^2)$ \emph{nonlinear} surfaces.)

For the case of integer input in the word RAM model, it is not difficult to obtain some polylogarithmic
speedup over $n^{2d/(d+1)}$ for Hopcroft's problem in $\R^d$, by using cuttings to reduce to
subproblems of polylogarithmic size, and solving these small subproblems by hashing
(modulo small primes) and bit packing (similar to known slightly subquadratic algorithms for integer 3SUM~\cite{BaranDP08}).  However, this only works for the version of the problem for
incidences, and not point-above-hyperplane pairs.

While it is difficult to prove good unconditional lower bounds on Hopcroft's problem
in general models of computation, one interesting direction (which has received
much recent attention in other areas of algorithms) is to prove
conditional lower bounds, based on the conjectured hardness of other problems, via fine-grained reductions.

For example, we can prove a conditional lower bound for Hopcroft's problem based on a problem known as affine degeneracy testing:


For a set of $m$ points in $\R^d$, the \textit{affine degeneracy testing problem} asks if there exist $d+1$ distinct points that lie on some  hyperplane.
The fastest algorithm known has $O(m^d)$ running time.
It has been conjectured that  the affine degeneracy testing problem cannot be solved faster than $O(m^{d-\eps})$ time, and lower bounds were known in some restricted computational models~\cite{Erickson99}.  The $d=2$ case (detecting 3 collinear points), in particular, is one of the standard 3SUM-hard problems in computational geometry~\cite{GajentaanO95}, believed to require near quadratic time.

We observe that affine degeneracy testing in 3D reduces to Hopcroft's problem in 5D (the observation is simple, but we are not aware of any explicit reference in the literature):

\begin{theorem}
  If Hopcroft's problem  for $n$ points and hyperplanes in $\R^5$ can be solved in $O(n^{3/2-\eps})$ time for some $\eps>0$, than the affine  degeneracy testing problem for $m$ points in $\R^3$ can be solved in $O(m^{3-\eps'})$ for some $\eps'>0$.
\end{theorem}
\begin{myproof}
  It will be more convenient to work with a colored version of the affine degeneracy testing problem: given $m$ points in $\R^3$ where each point is colored red, blue, green, or yellow, decide if there exists four points of different colors that lie on some hyperplane.  The original version reduces to the colored version, for example, by the standard (randomized or deterministic)  color-coding technique~\cite{AlonYZ95}, which increases running time only by polylogarithmic factors.

  Let $L_1$ (resp., $L_2$) be the set of $n = O(m^2)$ lines obtained by joining all red-blue (resp., green-yellow) pairs of points. We will use the Pl\"{u}cker transform to map the lines in $L_1$ to points and the lines in $L_2$ to hyperplanes in $5$-dimensional space.
  By standard facts about Pl\"{u}cker space \cite{ChazelleEGSS96}, two lines $\ell_1$ and $\ell_2$ are coplanar iff the point corresponding to $\ell_1$ lies on the hyperplane corresponding to $\ell_2$.
%
  Thus if we can solve 5D Hopcroft's problem in $O(n^{3/2-\eps})$ time, we can solve the 3D affine degeneracy testing problem in $O(m^{3-2\eps}\log^{O(1)}m)$ time.
\end{myproof}

Unfortunately the above $n^{3/2-\eps}$ conditional lower bound is not very tight, as we are only able to solve 5D Hopcroft's problem  in $O(n^{5/3})$ time. 
On the other hand, if 5D Hopcroft's problem turns out to have near linear algebraic decision tree complexity, then the above argument would imply better algebraic decision tree bounds for 3D affine linear degeneracy testing as well (which would lead to polylogarithmic speedup in the time complexity for this 3D problem).

The argument generalizes to give an $\Omega(n^{2d/(d+1)-\eps})$ conditional lower bound for Hopcroft's problem in $\binom{d+1}{(d+1)/2}-1$ dimensions for $d$ odd. However, the dimension of the transformed space is exponential in $d$.

\bibliographystyle{plain}
\bibliography{hopcroft}

\appendix

\end{document}